\documentclass[a4paper]{article}

\usepackage{natbib}
\usepackage[english]{babel}
\usepackage[utf8x]{inputenc}
\usepackage[T1]{fontenc}
\usepackage{amssymb}
\usepackage{enumitem}

\usepackage[toc,title,page]{appendix}

\newtheorem{property}{Property}

\usepackage[a4paper,top=3cm,bottom=2cm,left=3cm,right=3cm,marginparwidth=1.75cm]{geometry}

\usepackage{amsmath}
\usepackage{graphicx}
\usepackage[colorinlistoftodos]{todonotes}
\usepackage[colorlinks=true, allcolors=blue]{hyperref}
\usepackage{bm}
\usepackage{float}
\usepackage{rotating}
\usepackage{subcaption}

\newcommand{\fn}{\textrm{fines}}
\newcommand{\fb}{\textrm{fibers}}
\renewcommand{\v}{{\bf v}}
\newcommand{\bfx}{{\bf x}}

\newcommand{\bmth}{{\bm\theta}}

\usepackage[colorinlistoftodos]{todonotes}
\definecolor{darkolivegreen}{rgb}{0.33, 0.42, 0.18}
\definecolor{mediumpersianblue}{rgb}{0.0, 0.4, 0.65}
\definecolor{firebrick}{rgb}{0.7, 0.13, 0.13}

\title{Non-destructive methods for assessing tree fiber length distributions in standing trees}

\author{Sara Sj\"{o}stedt de Luna, Konrad Abramowicz, Natalya Pya Arnqvist \\
          \small Department of Mathematics and Mathematical Statistics, \\
         \small Ume\aa{}  University, 901 87 Ume\aa{}, Sweden\\[0.1cm]
          }
\date{}

\begin{document}
\maketitle

\begin{abstract}
One of the main concerns of silviculture and forest management focuses on finding fast, cost-efficient and non-destructive ways of measuring wood properties in standing trees. This paper presents an R package \verb+fiberLD+ that provides functions for estimating tree fiber length distributions in the standing tree based on increment core samples. The methods rely on increment core data measured by means of an optical fiber analyzer (OFA) or measured by microscopy. Increment core data analyzed by OFAs consist of the cell lengths of
both cut and uncut fibres (tracheids) and fines (such as ray parenchyma cells) without being able to identify which cells are cut or if they are fines or fibres. The microscopy measured data consist of the observed lengths of the uncut fibres in the increment core. A censored version of a mixture of the
fine and fiber length distributions is proposed to fit the OFA data, under distributional assumptions.
Two choices for the assumptions of the underlying density functions of the true fiber (fine) lenghts of those fibers (fines) that at least partially appear in the increment core are considered, such as the generalized gamma and the log normal densities. Maximum likelihood estimation is used for estimating the model parameters for both the OFA analyzed data and the microscopy measured data. 
\end{abstract}

\section[Introduction]{Introduction}
   There is an increased interest in improving the utilization of wood resources and the performance
of wood-based products. It has therefore become important to find ways of assessing
wood properties in growing trees, to serve tree breeding programmes and to evaluate silvicultural
methods. It is desirable that the procedures to measure wood properties in standing
trees are fast, cost-efficient and non-destructive (with respect to the tree). Rather recently, such methods have 
been proposed to measure fibre length and its distribution, 
see \cite{morling2003method, chen2016method,svensson2006estimation,svensson2007adjusting,svensson2010asymptotic}. In this paper we introduce 
an R-package that implements the methods proposed in these papers. We also develop numerical algorithms used to speed up computational time. 
   
\begin{figure}
\begin{center}
\includegraphics[width=1\linewidth]{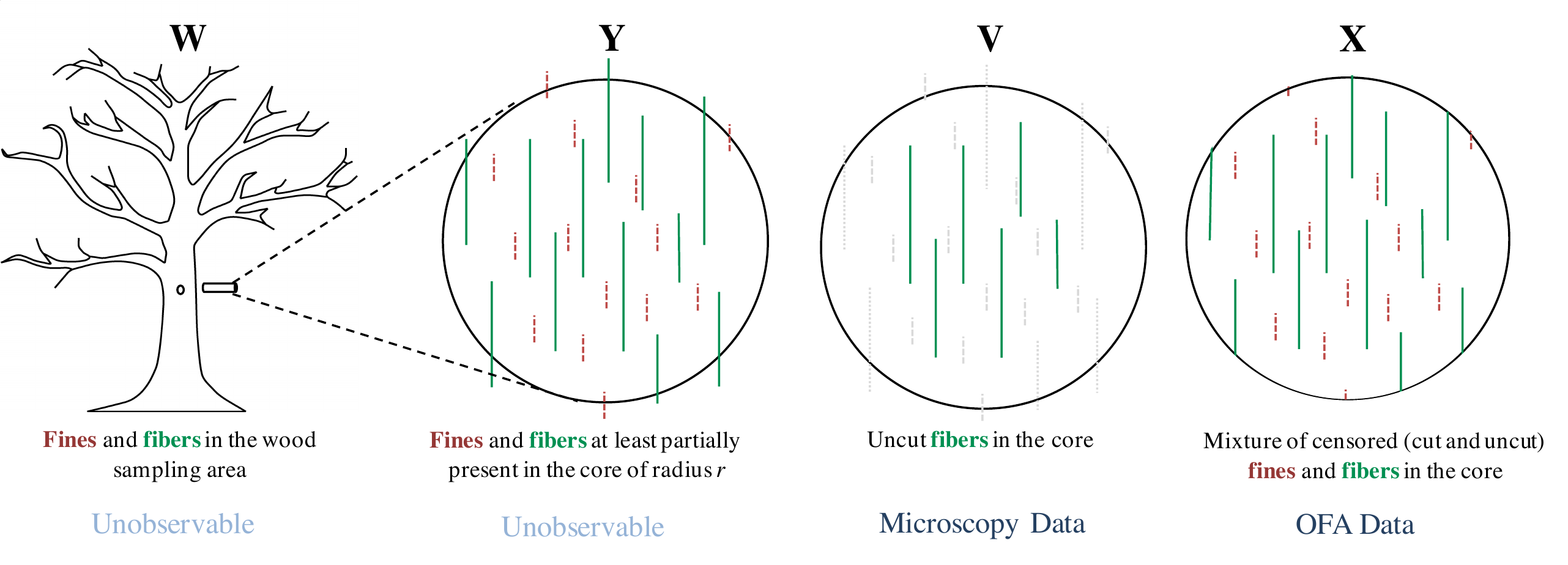}
\end{center}
\caption{Illustration of four different populations of cells from the same wood core measured by an optical fiber analyzer, OFA, (population X) or by microscopy (population V). The two unobservable populations are the cells in the wood sampling area of the tree (wood population, W) and the cells that are at least partially present in the core (core population, Y). The wood core is represented as a circle. Fibers and fines cells included in each population are marked by solid green and red dashed lines, respectively.}
\label{fig1}
\end{figure}

The methods rely on a commonly used fast sampling method, that is considered to be non-destructive, which is the
usage of increment cores with a diameter of 5 mm. An increment core is a cylindrical wood
sample taken in the tree with a special borer, see Fig. \ref{fig1}.

There are several reasons to why the fibre length distribution in an increment core is not the same as the fibre length
distribution in the standing tree. To begin with, the sample from an increment core will contain fibres cut once or twice, as well
as uncut fibres. This comes from the fact that the increment core is taken horizontally in the tree while fibres grow vertically,
see Fig. \ref{fig1}. All fibres exceeding the diameter of the core will certainly be cut. Secondly, the sample contains not only the fibres of
interest (tracheids) but also other types of cells such as ray parenchyma and ray tracheids that are much shorter; we will call them fines. The average
fibre length varies between 2 and 6mm for different softwood species, whereas the average fine length is about $10?20$ times
shorter \citep{ilvessalo1995fiber}. Another issue that needs to be taken into account is the length-bias problem arising from the fact
that longer cells are more likely to be sampled in an increment core.

In order to measure the individual lengths of the cells in a sample from an increment core, the cells are first totally separated by a chemical treatment \citep{franklin1945preparation}. The cell lengths  can be measured either in a microscope or by an automatic optical
fibre-analyser. In a microscope, it is possible to tell the difference between fines and fibres and
also whether or not a cell has been cut. However, a microscope analysis is not automated and
therefore the amount of cells that can be measured in practise is limited. An optical fibre-analyser (OFA)
can measure large samples automatically in a short time, but cannot automatically
distinguish fines from fibres and cannot tell if a cell is cut. This implies that
the observed lengths (measured by an OFA) come from a censored version of a mixture of the fine and the fibre length 
distributions in the tree, where the censoring mechanism caused by the
increment core is known. A
stochastic EM algorithm, that is capable of handling this lack of information, was proposed by \cite{svensson2006estimation} to estimate the fine and fibre length distributions under log normal assumptions. Consistency and asymptotic normality of the parameter estimates was confirmed by \cite{svensson2010asymptotic}. 

\cite{chen2016method}  applied the method proposed by \cite{svensson2006estimation} to estimate fibre length distributions in wood samples from Norwegian Spruce. The results were compared to estimated fibre length distributions based on microscopy measurements, where only the lengths of the uncut fibres in the sample were recorded. They derived the relationship between the length distribution of uncut fibres in the core and the one in the standing tree. 

The lognormal assumption used by \cite{svensson2006estimation}  enables a good EM algorithm to be proposed, but is somewhat limiting, since fibre length distributions may be skewed both to the left and to the right.  The generalized gamma distributions (GGD) is a more flexible family of distributions, of which the lognormal distribution is a limiting case. GGD were assumed for the microscopy measurements in \cite{chen2016method}. In the Appendix we derive the necessary equations for maximizing the observed log likelihood for both microscopy and OFA data. Good starting values are also suggested to avoid going into a sub-optimal maximum. Analytical derivatives used to speed up the optimization process are also given. The method is much faster than the EM algorithm and more flexible since it allows skewness in both directions. 

In this paper we introduce a package \verb+fiberLD+ that can estimate fine and fibre length distributions from increment core samples, from length measurements observed by an OFA. The package \verb+fiberLD+ is written in the R system for statistical computing \citep{R2020}, and is available from the Comprehensive R Archive Network at http://CRAN.R-project.org/. The EM algorithm of \cite{svensson2006estimation} for lognormally distributed data is implemented as well as a direct maximization of the log likelihood under lognormal or GGD distributional assumptions. Furthermore, estimation of fiber length distributions based on microscopy data under lognormal or GGD distributional assumptions is implemented. Estimation of some summary statistics of the length distributions and parameter estimates together with standard errors are also provided. 

Section \ref{inferencemethods} presents notation, formulas and inference methods in more detail. In Section \ref{implementation} the R-package \verb+fiberLD+ is presented. Detailed derivations of analytical derivatives of the log likelihood function used to find its maximum under GGD assumptions, etc is given in the Appendix.

%

\section{Inference methods}\label{inferencemethods}
In this section we provide an overview of the inference methods that are implemented in the proposed \verb+fiberLD+ R-package, used for the estimation of fine and fiber length distributions in a given wood sampling area of a standing tree, based on increment core samples. First, we introduce notation and densities for the random variables describing the length of cells in various populations (see Fig. \ref{fig1}): Let W denote the length of a cell (fine or fibre) in a standing tree. Further let Y denote
the true length of a cell that at least partially appears in the increment core, and X its
corresponding length seen in the increment core. Note that $X\leq Y$ since the cell might have
been cut. Finally, let $V$ denote the length of a (randomly chosen) uncut fibre in the core. 

The data observed from the microscopy measurements is a sample from population $V$ while the data from the optical fiber analyser is a  sample from population $X$.  
We are interested to estimate properties of the $W$ population, but have access to data from the distributions of $X$ and/or $V$. 
Because it is not possible to distinguish between fines and fibres when the
lengths are measured in OFAs, the resulting distributions of $Y,W$ and $X$ are
mixtures of some kind of a fine and a fibre length distribution.
Parametric assumptions are made on the distribution of $Y,$ 
which can be written as
\begin{equation}\label{pdfY}
  f_Y(y;\bmth)=\varepsilon f_{Y_\fn}(y;\bmth_\fn)+(1-\varepsilon)f_{Y_\fb}(y;\bmth_\fb),
\end{equation}
where $\varepsilon$ is the proportion of fines in the increment core, $0\leq \varepsilon \leq 1$ and $f_{Y_\fn}(y;\bmth_\fn)$ and $f_{Y_\fb}(y;\bmth_\fb)$ are the densities of the true lengths of fines and fibers, respectively, that at least partially appear in the core. Here $\bmth=(\varepsilon,\bmth_\fn, \bmth_\fb).$
The other distributions can be expressed in terms of these distributions, see Section \ref{subsection relationship}.


\subsection{Relationships between different cell length distributions}
\label{subsection relationship}
We now present the relationships between the introduced populations in Fig. \ref{fig1}. We start by describing the relation between the cell length distributions of Y and W. 
The distributions differ due to the length bias problem arising from the fact that longer cells are more likely to be sampled in the core.  Details and derivations are found in  \cite{svensson2006estimation}. 

\begin{property}[(Wood(W) and Core(Y)]
Assume that the core (of radius $r$) is randomly placed in the area of interest. 
Then the density function of the true fiber lengths in the standing tree, $W_\fb,$ satisfies
\begin{equation}\label{fWfibers}
  f_{W_\fb}(w;\bmth_\fb)=\frac{\pi r +2E(W_\fb)}{\pi r+2w}f_{Y_\fb}(w;\bmth_\fb), \;\; w>0,
\end{equation}
where 
\begin{equation}\label{EWfibers}
 E(W_\fb)=\frac{1}{2}\left[ \int_{0}^{\infty}\frac{f_{Y_\fb}(y;\bmth_\fb)}{\pi r+2y}dy\right]^{-1} -\frac{\pi r}{2},
\end{equation}
is the expected fiber length in the wood sampling area. The density function of the true fine lengths in the standing tree, $f_{W_\fn}(w;\bmth_\fn),$ and the expected value $E(W_\fn)$ are defined analogously by replacing "fibers"  by "fines" in \eqref{fWfibers} and \eqref{EWfibers}.
\end{property}

We further have that the density function of the cell length distribution in the standing tree is the mixture density
$$
  f_W(w;\bmth)=\tilde{\varepsilon} f_{W_\fn}(w;\bmth_\fn)+(1-\tilde{\varepsilon})f_{W_\fb}(w;\bmth_\fb), \;\; w>0,
$$
where
$\tilde{\varepsilon} $ is the proportion of fines in the staning tree, satisfying
\begin{equation}\label{tilde.eps}
  \tilde{\varepsilon} =\varepsilon\left[\frac{\pi r+2E(W)}{\pi r+2E(W_\fn)} \right],
\end{equation}
and the expected cell length in the tree, $E(W),$ corresponds to
\begin{equation}\label{EW}
 E(W)= \frac{2E(W_\fn)E(W_\fb)+\varepsilon \pi r E(W_\fn)+(1-\varepsilon)\pi rE(W_\fb)}{2\left\{\varepsilon E(W_\fb)+(1-\varepsilon)E(W_\fn)\right\} +\pi r}.
\end{equation}

Hence, it follows that if we know the cell length distributions in population $Y$, it is sufficient for finding analytical expressions describing the cell length distributions in population $W$. 

To incorporate the censoring mechanism, we need the probability $p_{UC}(y)$ for a cell of true length $y$ to be uncut in a core of radius $r$. Under the assumption that the core is randomly placed in the region of interest, and that cells are randomly packed,  \cite{svensson2006estimation} show that this probability can be expressed as:
$$
p_{UC}(y)=\frac{1}{t(y)}\left(2r^2\mathrm{arcsin}\left(\frac{\sqrt{4r^2-y^2}}{2r}\right)-\frac{y}{2}\sqrt{4r^2-y^2}\right),\qquad 0<y \leq 2r,
$$
with $t(y)=\pi r^2+ 2 r y$. Note that $p_{UC}(y)=0$ if $y > 2r$ since a cell longer than the diameter of the core necessarily is cut.
Using the above we have the following property that describes how the fiber length density of the uncut fibers in the core (population $V=V_\fb$) can be expressed in terms of the fiber length density $f_{Y_\fb}(y;\bmth_\fb),$ for proof see \citep{chen2016method}. 
\begin{property}[Microscopy(V) and Core(Y)]
{
\noindent  Assume that the core (of radius $r$) is randomly placed in the area of interest. Then the distribution of the lengths of uncut fibers in the core, described by density $f_V(\cdot)=f_{V_\fb}(\cdot) $ satisfies
\begin{equation}
f_{V_\fb}(v;\bmth_\fb)=\frac{1}{k_{\bmth_\fb}}f_{Y_\fb}(v;\bmth_\fb) p_{UC}(v), \qquad 0<v<2r, 
\label{eq:VY}
\end{equation}
with $k_{\bmth_\fb}=\int_0^{2r}f_{Y_\fb}(v;\bmth_\fb) p_{UC}(v)\mathrm{d}v$. 
}
\end{property}

A sample from the distribution described by $f_{V_\fb}(\cdot) $ can be observed by microscopy, measuring the lengths of uncut fibers in the core. Additionally, the expression for $p_{UC}(\cdot)$ together with insight in the censoring mechanism induced by the core, provides us with the relationship between the densities of cell lengths on the Y and X scale. Therefore we can express the densities for the observed sample from population X in terms of the distribution of cells in the core \citep[c.f., ][]{svensson2006estimation}.

\begin{property}[OFA(X) and Core(Y)]
{
\noindent Assume that the core (of radius $r$) is randomly placed in the area of interest. Then the density function of the observable lengths of (cut and uncut) fines in the core, 
 $f_{X_\fn}(\cdot)$, satisfies,
\begin{equation}
f_{X_\fn}(x;\bmth_\fn) = p_{UC}(x)f_{Y_\fn}(x;\bmth_\fn)+\int_x^\infty k_{X|Y}(x|y)f_{Y_\fn}(y;\bmth_\fn)dy,\qquad 0<x<2r,\\
\end{equation}
\noindent where 
$$k_{X|Y}(x|y)=\frac{8r^2-3x^2+yx}{t(y)\sqrt{4r^2-x^2}}.$$ 

The density function of the observable fibre lengths in the core, $f_{X_\fb}(x;\bmth_\fb)$, is defined analogously.} 
\end{property}

The density function of the cell length distribution in the core
is the mixture density,
\begin{equation}\label{pdfX}
  f_X(x;\bmth)=\varepsilon f_{X_\fn}(x;\bmth_\fn)+(1-\varepsilon)f_{X_\fb}(x;\bmth_\fb),\qquad 0<x<2r,
\end{equation}
where $\varepsilon$ is the proportion of fines in the increment core. 

The above properties are the key to inference for the observable data. Under parametric assumptions on the densities of the core population $Y$, we can express the observable likelihood functions for the observable populations that we can take samples from ($V$ and $X$). Further, after estimating the parameters of the densities on the $Y$ scale, we can transform the densities to the $W$ scale by using the relations describe above and plugging in the estimated parameters $\hat{\theta}$, obtaining full information about the 
 fine and fiber lenght distributions in the wood sampling area of the standing tree. 

\subsection{Estimation from Microscopy Data}
\label{subSec:estimateMic}
Suppose that we have observed a random sample of $n$ uncut fibers from the core, with their lengths, $\v=(v_1,\ldots,v_n),$ being measured manually using microscope. 
We further assume that the true lengths of fibers that at least partially appear in the core follow a generalized gamma distribution (GGD), i.e. the density $f_{Y_\fb}(y;\boldsymbol{\theta}_\fb)$ satisfies (\ref{GGDfb}), see also \eqref{GGDthetafb}.
Then the observed log likelihood is obtained by 
\begin{equation}
\label{logLmicro}
\log L(\boldsymbol{\theta}_\fb;\v)=\sum_{i=1}^n \log f_{V_\fb}(v_i;\boldsymbol{\theta}_\fb) \propto \sum_{i=1}^n \log f_{Y_\fb}(v;\bmth_\fb) - n\log{k_{\bmth_\fb}},
\end{equation}
cf. equation \eqref{eq:VY}. The parameters are estimated by maximizing \eqref{logLmicro} via a direct optimization algorithm. In our proposed R-package \verb+fiberLD+ we use the existing  R-function \verb+optim+, which performs general-purpose optimization based on Nelder--Mead, quasi-Newton and conjugate-gradient algorithms. The \verb+optim+ function may use numerical or analytical derivatives, and \verb+fiberLD+ allows both alternatives. Multiple initial points are used to facilitate global maximum discovery. 

\subsection{Estimation from OFA Data}
Suppose that we have observed the lengths of $n$ cells in the core, $\bfx=(x_1,\ldots,x_n),$ measured by an optical fiber analyzer, the cells being a mixture of cut and uncut fines and fibers. We call this OFA data. By equation \eqref{pdfX}, the observed log likelihood of the OFA data corresponds to
\begin{equation}
\log{L(\boldsymbol{\theta};\bfx)}=\sum_{i=1}^n \log
{f_X(x_i;\boldsymbol{\theta})}=
\sum_{i=1}^{n}\log{\left[\varepsilon f_{X_\fn}(x_i; \boldsymbol{\theta_\fn})+ (1-\varepsilon) f_{X_\fb}(x_i,\boldsymbol{\theta_\fb}) \right]},
\label{LfX}
\end{equation}
where $\boldsymbol{\theta}=(\boldsymbol{\theta_\fn},\boldsymbol{\theta_\fb}, \varepsilon)$. 
We propose two alternative estimation approaches which differ in the distributional assumptions on the mixture components and the estimation methods. Both methods target on maximizing the \textit{observed log-likelihood of the OFA data} in \eqref{LfX}.

\subsection*{Alternative I - Generalized gamma distributions}
{
Assume that the densities of the true lenghts of fibers and fines that at least partially appear in the core follow generalized gamma distributions, such that $f_{Y_\fn}(y;\bmth_\fn)$ and $f_{Y_\fb}(y;\bmth_\fb)$ satisfy \eqref{GGDfn} - \eqref{GGDthetafb}.
By Property 3, we have an analytical expression of the observed log likelihood \eqref{LfX} of the OFA data in terms of the assumed generalized gamma densities. 
The parameters $(\boldsymbol{\theta_\fn},\boldsymbol{\theta_\fb},\varepsilon)$ are estimated by maximizing the {observed log-likelihood} \eqref{LfX} via the direct optimization algorithm described in Section 
\ref{subSec:estimateMic}. The analytical gradient and Hessian of  \eqref{LfX} are derived in the Appendix. 
}
\subsection*{Alternative II - Lognormal distributions}
{
Assume that the densities of the true lenghts of fibers and fines that at least partially appear in the core  follow lognormal distributions such that that $f_{Y_\fn}(y;\bmth_\fn)$ and $f_{Y_\fb}(y;\bmth_\fb)$ satisfy \eqref{LogNfn} - \eqref{LogNtheta}.
Note that the lognormal distribution is a limiting distribution of the generalized gamma distribution and a member of the exponential family of distributions. Under these assumptions we propose two ways of estimating the parameter $\boldsymbol{\theta}=(\boldsymbol{\theta_\fn},\boldsymbol{\theta_\fb},\varepsilon)$. The first method maximizes the observed log-llkelihoood \eqref{LfX} via a direct optimization algorithm. The second method uses a stochastic version of an EM algorithm proposed by \cite{svensson2006estimation} to maximize the observed log likelihood \eqref{LfX}. For the latter method weak consistency and asymptotic normality of the estimator is derived \cite{svensson2010asymptotic} and therefore we propose it as a suitable alternative to the first numerical maximization algorithm. For details of the algorithm, we refer to \cite{svensson2006estimation}.
}

\section{Implementation}\label{implementation}
    The methods for estimating tree fiber length distributions in the standing tree based on increment core samples presented in the previous sections have been implemented in 
    our R package \verb+fiberLD+. Two types of data can be used with the package, OFA data and Microscopy data. 
    The code below loads the package and the simulated data used in the examples.
\begin{verbatim}
 > library(fiberLD)
 > data("cell.length", package="fiberLD")
 > data("microscopy", package="fiberLD")
\end{verbatim}

    The key functions of the package with short descriptions are listed in Table \ref{table1}. The main routine \verb"fled" is complemented with other routines that help to summarize, print and plot 
the results of the estimation procedure. The S3 method \verb"summary" summarizes the optimization results and does post processing of the model parameters estimates
by returning the summary statistics (the mean, standard deviation, skewness, kurtosis and the corresponding standard errors) of the fiber and fine lengths in the standing tree (population $W$), 
the expected value of the cell lengths, $E[W]$, the proportion of fines in the standing tree, $\tilde{\epsilon}$, and its standard error. The S3 method \verb"plot" creates the estimated density of the mixture model, $f_W(\cdot)$, 
together with the histogram of the data and the estimated densities of the fiber and fine lengths, $f_{W_\fb}(\cdot)$ and $f_{W_\fn}(\cdot)$.
  The following subsections are addressed to each of the key functions. 

    \begin{table}\caption{The main functions of the R package fiberLD} \label{table1}
     \centering
         \begin{tabular}{l|l}
             \hline
             Function  & Description\\
              \hline
               \verb"fled" & the main function to estimate cell length distributions \\
                 &  \\
               \verb"print.fled" & printing the basic model estimation information  \\
                &  \\
                \verb"plot.fled" & plotting various estimated fiber/fine length densities \\
                  &  \\
                \verb"summary.fled" & extracting the model estimation results \\
                 &  \\
                 \verb"dx.fibers" & fiber/fine length density evaluation on the $X,$ $Y,$ and \\
                 \verb"dy.fibers" & $W$  scales based on OFA data\\
                 \verb"dw.fibers" &  \\
                 &  \\
                 \verb"dx.fibers.micro" & fiber length density evaluation on the $X,$ $Y,$ and \\
                 \verb"dy.fibers.micro" & $W$  scales based on microscopy data\\
                 \verb"dw.fibers.micro" &   \\
                   &  \\
                 \verb"dx.mixture" & mixture density evaluation on the $X,$ $Y,$ and \\
                 \verb"dy.mixture" & $W$  scales \\
                 \verb"dw.mixture" &  \\
            \hline
       \end{tabular}
  \end{table}

\subsection{fled function}

 The function \verb"fled"  is the main routine which implements parameter estimation for both the OFA analyzed data and the microscopy measured data. 
 The function offers two choices for the underlying density functions $f_{Y_\fn}(y,\bmth_\fn)$ and $f_{Y_\fb}(y,\bmth_\fb)$; the
 generalized gamma and lognormal densities. The parameters $\boldsymbol{\theta}=(\boldsymbol{\theta_\fn},\boldsymbol{\theta_\fb},\varepsilon)$ are estimated by log likelihood maximization as described in Section 2. 
 As an example, the OFA data with underlying generalized gamma density functions can be analyzed by calling

 \begin{verbatim}
 > d1<-fled(data=dat,data.type="ofa",r=6,model="ggamma",method="ML")
\end{verbatim}
 where the \verb"data" argument of the \verb"fled" function is a numeric vector of cell lengths from an increment core,
\verb"data.type" denotes the type of data supplied which can be either \verb"`ofa'" (default) or \verb"`microscopy'". The radius of the increment core is denoted by \verb"r", \verb"model" indicates the distribution of the true fiber (fine) lengths that at least partially appear in the increment core; \verb"model=`ggamma'" corresponds to generalized gamma distributions, and log normal distributions are indicated by \verb"model=`lognorm'". The default method for parameter estimation corresponds to direct maximization of the log likelihood indicated by the argument \verb"method=`ML'". In addition a stochastic version of the expectation-maximization method is provided to fit the log normal model to the increment core data analyzed by OFAs. This is indicated by setting \verb"method=`SEM'".

The following line demonstrates how to perform parameter estimation of the log normal-based model by a stochastic version of the EM algorithm.
 \begin{verbatim}
 > d2<-fled(data=dat,data.type="ofa",model="lognorm",method="SEM")
\end{verbatim}

The routine \verb"fled" calls the optimization functions \verb"optim()" or \verb"nlm()" to maximize the log likelihood, with the possibility to use a supplied gradient function. The argument \verb"optimizer" specifies a numerical optimization method used to
minimize minus the log likelihood function of the observed data: \verb"`optim'", \verb"`nlm'" or \verb"`nlm.fd'" (\verb"nlm" is based on finite-difference approximation of the derivatives).
If \verb"`optim'" is called then two other components can be supplied to the argument \verb"'optimizer'".
The second component specifies the numerical method to be used in \verb"optim" (\verb"`L-BFGS-B'", \verb"`BFGS'", \verb"`CG'", \verb"`Nelder-Mead'" or \verb"`SANN'"). The third element of \verb"optimizer" indicates whether the finite difference approximation should be used (\verb"`fd'") or the analytical gradient (\verb"`grad'") for the \verb"`L-BFGS-B'", \verb"`BFGS'", \verb"`CG'" methods.

The default is \verb"optimizer=c(`optim',`L-BFGS-B',`grad')". When using maximum likelihood estimation for the censored mixture of  generalized gamma distributions, a quasi-Newton method with bound constraints that allow to set lower and upper limits (\verb"lower" and \verb"upper") for each parameter (\verb"`L-BFGS-B'") is recommended. Poor initial values may also result
in convergence problems. We suggest to use our built in routine for estimating initial values of the model parameters by assuming an uncensored version of the mixture of
generalized gamma (log normal) distributions. Still, the function \verb"fled" allows to set starting values of the parameters by supplying a numerical vector to the 
argument \verb"parStart". The order of the parameter values should be the same as given in Section \ref{inferencemethods}. 
The code below shows how to set lower and upper bounds for the \verb"`L-BFGS-B'" optimization procedure using the \verb"cell.length" OFA data.
\begin{verbatim}
 > d3 <- fled(data=cell.length, model="ggamma", r=6, 
        optimizer=c("optim","L-BFGS-B","grad"),
        lower=c(.12,1e-3,.05,rep(.3,4)), upper=c(.5,2,rep(7,5)))
\end{verbatim}

By typing \verb"d3" or \verb"print(d3)" the short-form model summary is printed.
\begin{verbatim}
 > d3
 Increment core data (all fiber and fine lengths in the core)

Model: Generalized gamma

Model parameters:
 0.2978 0.001 0.2921 5.252 2.001 2.822 2.224 

'-'Loglik = 3625.888  n = 3000
\end{verbatim}
The print method displays the type of data used, the model distribution assumed, and the estimated model parameters in the order
$(\varepsilon,b_1, d_1, k_1,b_2, d_2, k_2)$, where 1 stands for $\fn$ and 2 for $\fb.$ The value of the minus log likelihood and the 
number of observations used are reported at the end. The estimated values of the parameters of the fiber length distribution, for instance,
are $b_2=2.001,~~ d_2=2.822,~~ k_2=2.224.$

In addition, there is a possibility to fix some parameters of the generalized gamma mixture model at their 
pre-specified values supplied via the argument \verb"parStart". The additional TRUE/FALSE 
vector \verb"fixed" is added to indicate which parameters have to be fixed. 
The default is \verb"fixed=NULL". In case of fixing parameters, the positive values in \verb"parStart" for non-fixed parameters are treated as starting values 
for the optimiser, the negative or zero values indicate that no starting values are assumed. Fixing parameter values currently works only with \verb"optim".
Below is an example when fixing two parameters $d_1=d_2=1$ and setting initial values for the rest of the parameters at $\varepsilon=.5, ~~ b_1=.01, ~~k_1=1,~~b_2=2, ~~k_2=1.$ 
\begin{verbatim}
 > d4<-fled(data=dat,model="ggamma",parStart=c(.5,.01,1,1,2,1,1),
         fixed=c(FALSE,FALSE,TRUE,FALSE,FALSE,TRUE,FALSE)) 
\end{verbatim}
It should be noted that fixing parameters should be handled with care as it may lead to instability of the optim method.

\subsection{Plot method}

\begin{figure}[H]
   \begin{center}
         \includegraphics[height=10cm,width=11cm]{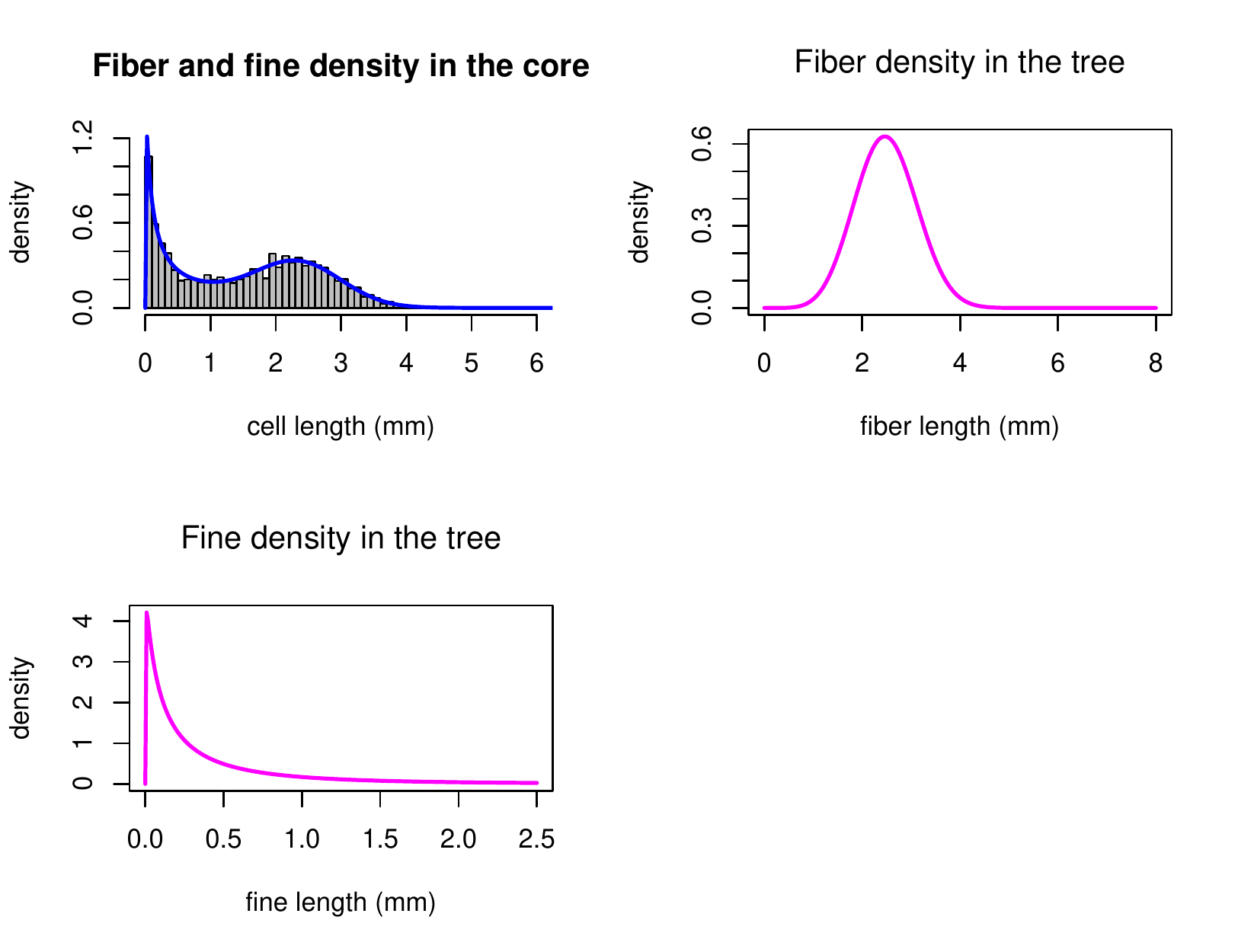}
   \end{center}
   \caption{Estimated density plots of model d3.}  \label{fig-ofa-ggamma}
\end{figure}

 When the \verb"fled" object is passed to the \verb"plot()" function, the plot method produces the estimated density plots.  
For OFA data (\verb"`ofa'"), the default plotting creates a histogram of 
the given data together with the estimated density of the mixture model, $f_X(\cdot),$ and two separate plots of the estimated fiber and fine lengths 
densities in the standing tree, $f_{W_\fb}(\cdot)$ and $f_{W_\fn}(\cdot).$ This is illustrated on the \verb"cell.length" example from the previous subsection (see Fig. \ref{fig-ofa-ggamma}).
\begin{verbatim}
 > plot(d3)
\end{verbatim}

\begin{figure}[H]
    \begin{center}
            \includegraphics[height=7cm,width=8cm]{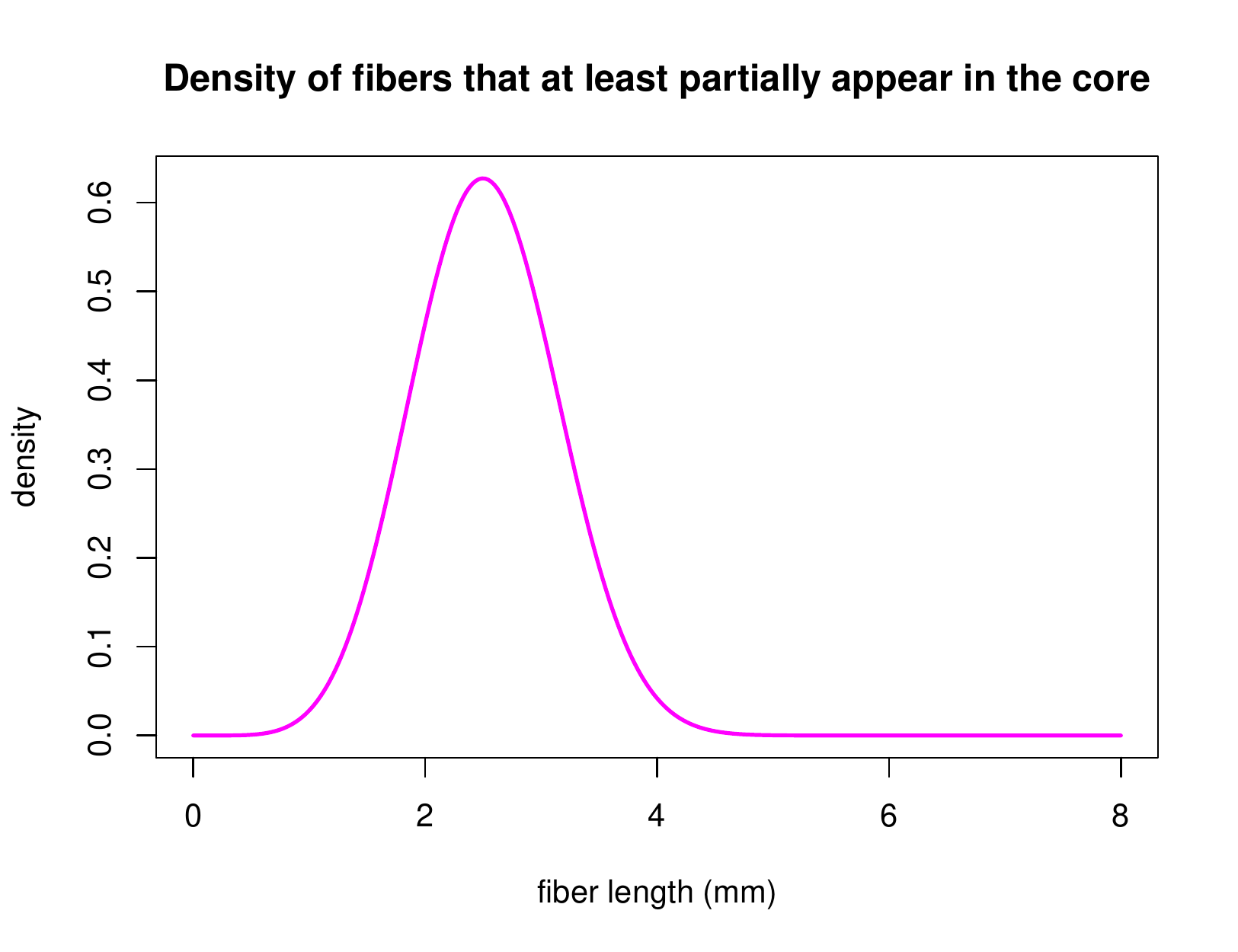}
    \end{center}   
    \caption{Density plot of the fiber length of those fibers that at least partially appear in the increment core.} \label{fig-ofa-ggamma-2}
\end{figure}

 It is possible to select one single plot to print by using the argument \verb"select", which can be set to either $1$ (the estimated 
 density of the mixture model and the histogram), $2$ (the estimated fiber length density) or $3$ (the corresponding fine length density). The default value is \verb"select=NULL."
  In addition by using the argument \verb"density.scale", the plot method provides an option to define the scale on which the fiber (fine) length densities should be plotted.
 \verb"density.scale" can be set to one of the three options: \verb"`tree'" (default) plots the estimated densities of the fiber (fine) lengths in the tree ($W$ scale),
 \verb"`uncut.core'" plots the densities of the cell (fiber or fine) lengths of those cells that at least partially appear in the increment core ($Y$ scale), 
 \verb"`core'" plots the densities of the observed (cut or uncut) cell lengths in the increment core ($X$ scale). The following code demonstrates how to plot the fiber length density of those fibers that
 at least partially appear in the increment core. The result is displayed as Fig. \ref{fig-ofa-ggamma-2}.
 \begin{verbatim}
 >  plot(d3,select=2,density.scale='uncut.core')
 \end{verbatim}
 
 By specifying  the \verb"rvec" argument the plot method allows modifications of the values of the cell lengths used for calculating estimates of the densities. There are also 
 arguments for specifying plot labels (\verb"xlab" and \verb"ylab" for axes labels, and \verb"main" for a title), for defining the color used for density plotting (\verb"col"),
 and the line width (\verb"lwd").
 
\begin{figure}[H]
     \begin{center}
           \includegraphics[height=10cm,width=11cm]{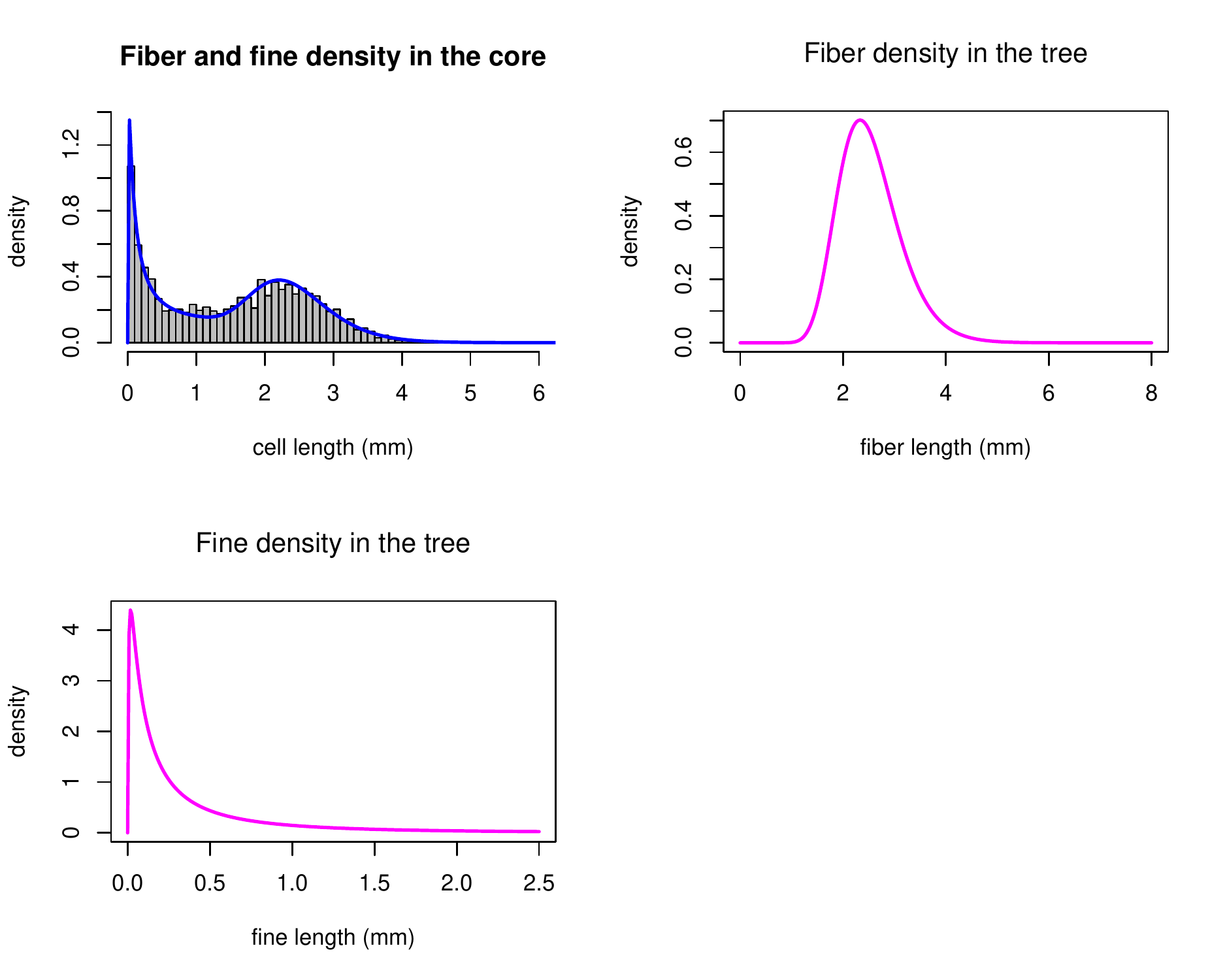}
      \end{center}   
   \caption{Estimated density plots of model d5.} \label{fig-d5}           
\end{figure}

 The lines below show how to fit the same data, \verb"cell.length",  and plot the results under the assumption that the fiber and fine length distributions that at 
 least partially appear in the core follow lognormal distributions. The plots are presented in Fig. \ref{fig-d5}.  The estimated model parameters are in the order
$(\varepsilon,\mu_1, \sigma_1,\mu_2, \sigma_2)$.
 \begin{verbatim}
 > d5 <- fled(data=cell.length,model="lognorm",r=6)
 > d5

 Increment core data (all fiber and fine lengths in the core)

 Model: Log normal

 Model parameters:
  0.2928 -1.58 1.555 0.9152 0.2382 

 '-'Loglik = 3645.569  n = 3000

> plot(d5)
 \end{verbatim}
 
\subsection{Summary method}

The summary method provides a list of summary information for a \verb"fled" object such as a table of the estimated model parameters, tables of summary statistics for fiber and
fine lengths in the standing tree, expected value of the cell length in the standing tree, proportion of fines in the standing tree and its standard error. The summary method
for the \verb"cell.length" example based on generalized gamma distributions produces:
\begin{verbatim}
> summary(d3)

Increment core data (all fiber and fine lengths in the core)

Model: Generalized gamma   Method: ML

Model parameters:
            b_fines  d_fines  k_fines b_fibers d_fibers k_fibers   eps
Estimate   0.001000 0.292066 5.251900 2.001418 2.822360 2.223642 0.298
Std. Error 0.002736 0.064486 2.159753 0.418737 0.581249 0.875709 0.015

Summary statistics for FIBER lengths in the standing tree:
              Mean Std.dev. Skewness Kurtosis
Estimate   2.48880  0.62405   0.1011    2.872
Std. Error 0.02671  0.02078   0.1077    0.070

Summary statistics for FINE lengths in the standing tree:
              Mean Std.dev. Skewness Kurtosis
Estimate   0.49992   0.8459   5.2904   52.167
Std. Error 0.05343   0.1114   0.4094    6.835

Proportion of fines in the standing tree: 0.34 (Std.error = 0.015)

'-'Loglik = 3625.888   Sample size: n = 3000

Convergence: Successful completion
 \end{verbatim}

The routine summary also prints the information about the type of data, the model and method used, the number of observations and the values of the minus log likelihood
of the fitted model. The last line indicates why the optimization algorithm terminated. To compare the generalized gamma based model (\verb"d3") with the log normal one (\verb"d5"),
the summary function is called for \verb"d5".
\begin{verbatim}
> summary(d5)

Increment core data (all fiber and fine lengths in the core)

Model: Log normal   Method: ML

Model parameters:
            mu_fines sig_fines mu_fibers sig_fibers   eps
Estimate   -1.579890  1.554697  0.915158   0.238151 0.293
Std. Error  0.089306  0.052468  0.008886   0.007867 0.014

Summary statistics for FIBER lengths in the standing tree:
              Mean Std.dev. Skewness Kurtosis
Estimate   2.53789  0.61006  0.73363    3.972
Std. Error 0.02162  0.02006  0.02552    0.069

Summary statistics for FINE lengths in the standing tree:
              Mean Std.dev. Skewness Kurtosis
Estimate   0.53531   1.0605   6.5411   67.792
Std. Error 0.05582   0.1032   0.2847    7.632

Proportion of fines in the standing tree: 0.33 (Std.error = 0.014)

'-'Loglik = 3645.569   Sample size: n = 3000

Convergence: Successful completion
 \end{verbatim}

\subsection{Density  functions}

\begin{figure}[H]
  \begin{center}
           \includegraphics[height=7cm,width=7cm]{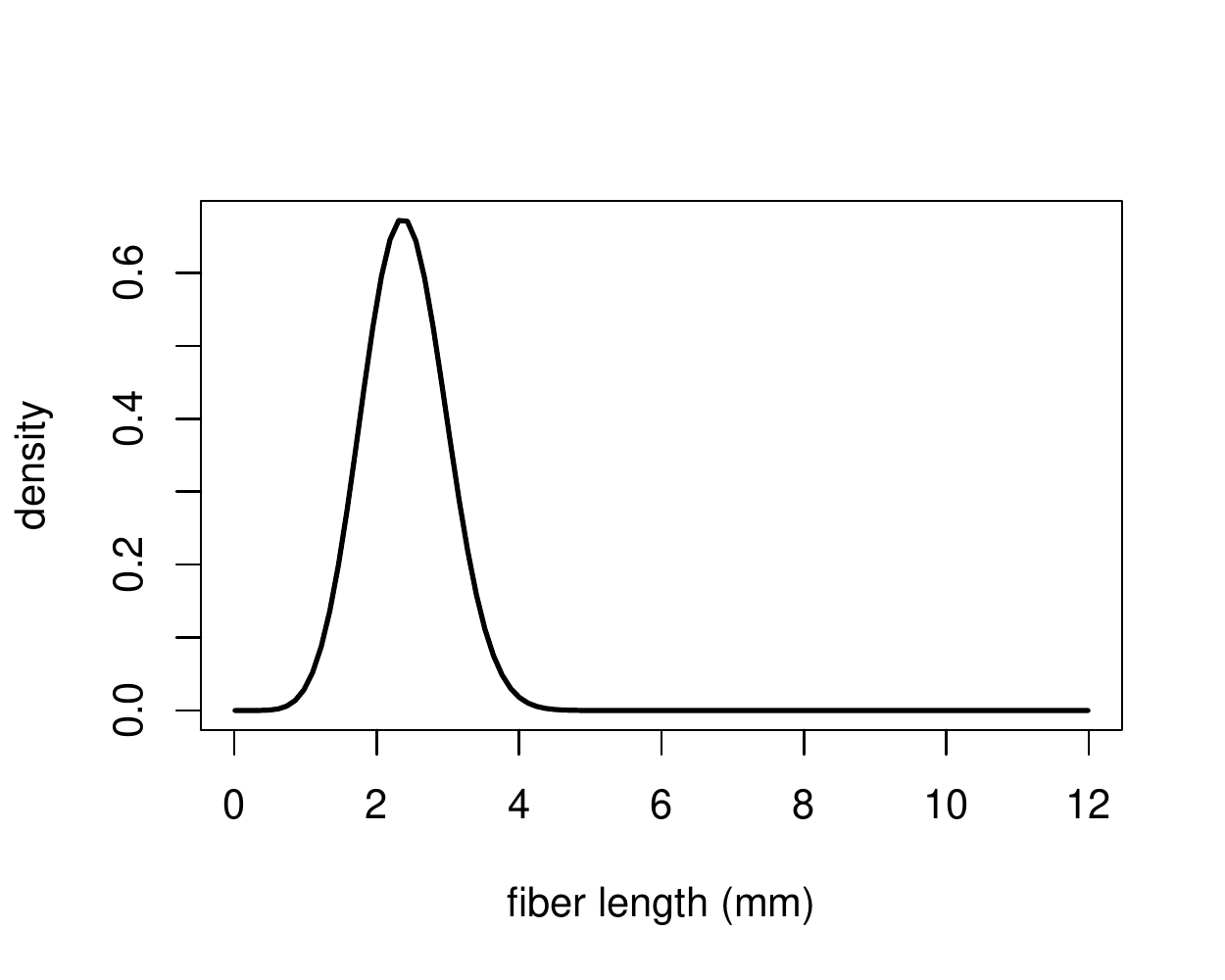}
  \end{center}           
  \caption{Density of the true fiber length in the standing tree.}\label{fig-fiber-density}
\end{figure}

The \verb+fiberLD+ package also contains functions that calculate the densities of the fiber lengths for the OFA and the microscopy analyzed data and 
of the mixture model on the three different scales: as observed in the increment core, as the true lengths of the fibers that at least partially appear in the increment core
and as the true fiber lengths in the standing tree.  The functions \verb"dx.fibers", \verb"dy.fibers" and \verb"dw.fibers" calculate values of the fiber length density 
function on the above mentioned three scales correspondingly. For example, to get the density values of the true fiber lengths in the standing tree, $f_{W_\fb}(\cdot),$ and plot them (see Fig. \ref{fig-fiber-density}),
the following code can be called. 
\begin{verbatim}
>  x <- seq(.01, 2*r -.01,length=100)
>  f <- dw.fibers(x, par=c(1.8,2.7,2.6), r=2.5)
>  plot(x,f,type="l",lwd=2,ylab="density",xlab="fiber length (mm)")
\end{verbatim}
 
The same functions can be used to plot fine lengths densities. An example on how to calculate fine lengths density values based on the log 
normal distribution is given below.
\begin{verbatim}
> par.fines <- c(-2, .5)
> x <- seq(.1, 1.5,length=5)
> f1.fines <- dy.fibers(x, par.fines, model="lognorm")
> f1.fines
[1] 6.643761e+00 9.882040e-02 1.805197e-03 7.317069e-05 5.011470e-06
\end{verbatim}

\begin{figure}[H]
  \begin{center}
           \includegraphics[height=7cm,width=7cm]{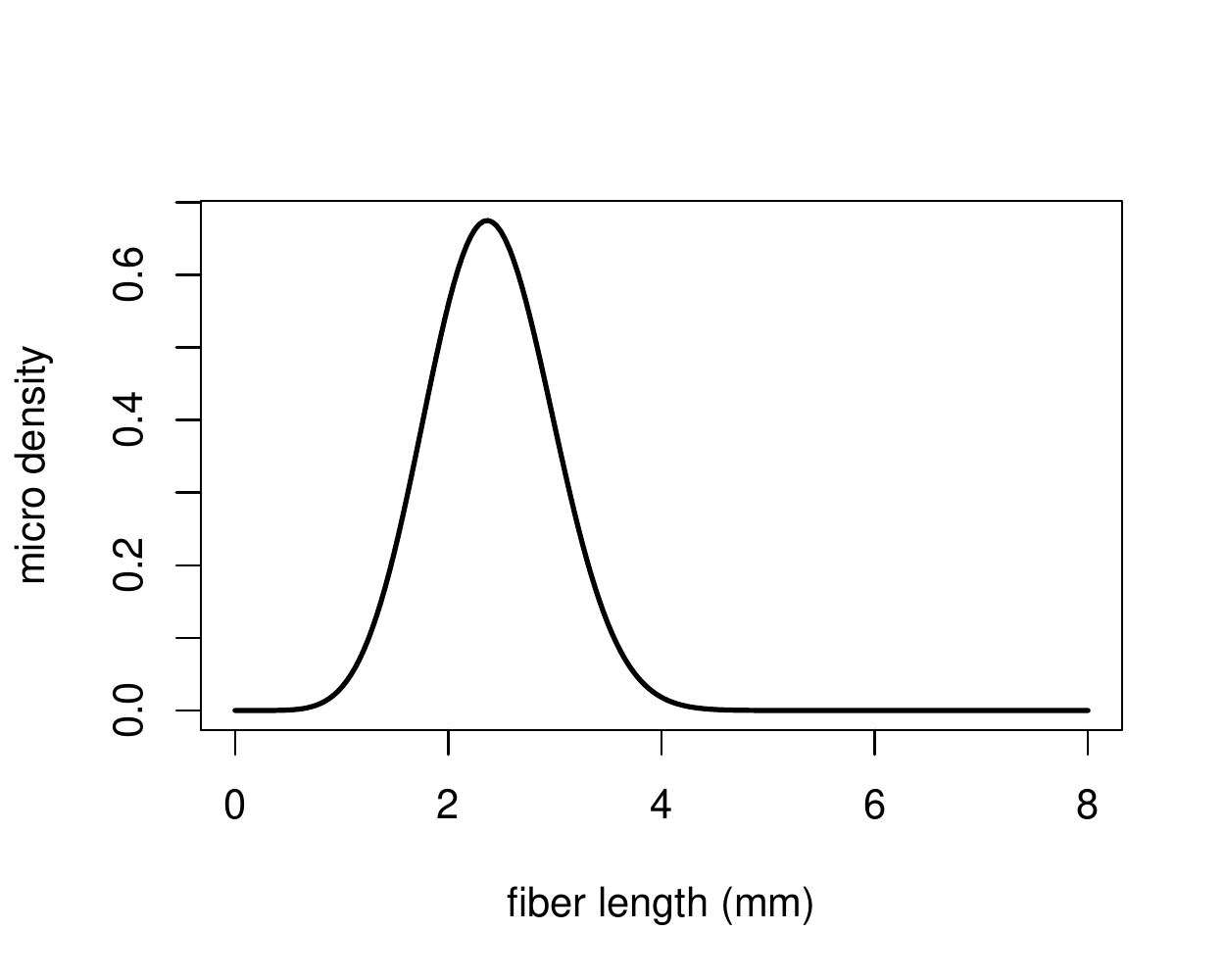}
  \end{center}           
  \caption{Density of the fiber length in the tree that goes beyond the length of the increment diameter.}\label{fig-fiber-micro-density}
\end{figure}

To obtain the density values of the fiber lengths based on microscopy data the routines \verb"dx.fibers.micro", \verb"dy.fibers.micro" and \verb"dw.fibers.micro" can be used.
A simple function-call that calculates the density values of the true fiber lengths of those fibers that at least partially appear in the increment core can be made as follows:
\begin{verbatim}
> f2 <- dy.fibers.micro(x=seq(0, 5,length=7), par=c(1.8,2.7,2.6))
> f2
[1] 0.0000000000 0.0089776518 0.2929873243 0.6689186996 0.2184220375
[6] 0.0106544667 0.0000692969
\end{verbatim}

The following code shows how to get the values of the fiber length density function in the tree that goes beyond the length of the increment core diameter. The result is displayed as Fig. \ref{fig-fiber-micro-density}.
\begin{verbatim}
> w <- seq(0,8,length=200)
> f3 <- dw.fibers.micro(w, par=c(1.8,2.7,2.6), r=2.5)
> plot(w,f3,type="l",lwd=2,ylab="micro density",xlab="fiber length (mm)")
\end{verbatim}

Finally, the mixture density functions of the cell lengths can be analyzed on three different scales using the functions \verb"dx.mixture", \verb"dy.mixture" and \verb"dw.mixture".
For example, the following code gets values of the mixture density of the cell lengths as observed in the increment core.
\begin{verbatim}
> d <- fled(data=cell.length,model="lognorm",r=6)
> x <- seq(0, 8,length=5)
>  f4 <- dx.mixture(x=x, par=d$par,r=6, model="lognorm")
>  f4
[1] 0.0000001000 0.3562456545 0.0216679677 0.0007470903 0.0002581674
\end{verbatim}
The other two functions, \verb"dy.mixture" and \verb"dw.mixture", can be used in a similar way.

\section{Using fiberLD with microscopy data}
 
\begin{figure}[H]
  \begin{center}
           \includegraphics[height=7cm,width=7cm]{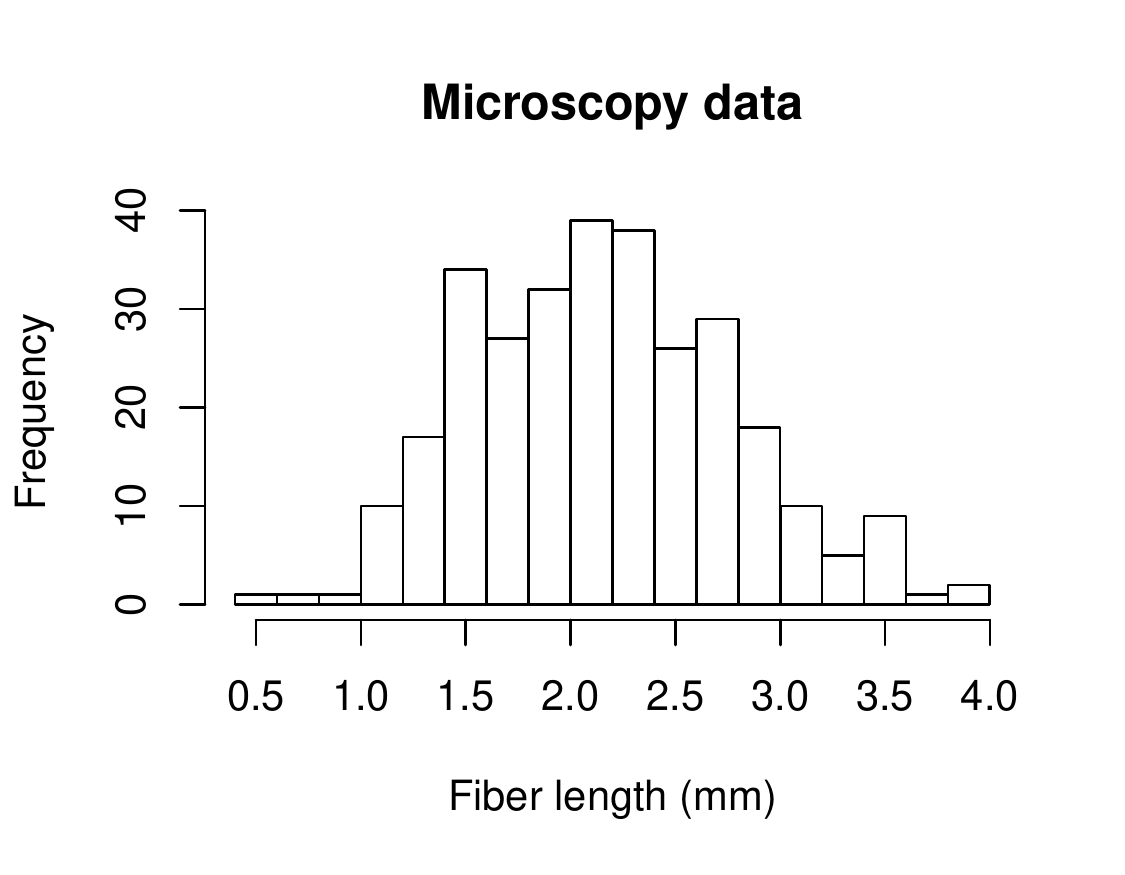}
   \end{center}           
   \caption{Histogram of simulated microscopy data.}\label{fig-hist-micro}
\end{figure}

In this section, we show how to apply the methods and functions of the \verb"fiberLD" package to microscopy data. The main routines are demonstrated on the simulated
dataset, \verb"microscopy", that is included in the package. \verb"microscopy" is a vector of 300 uncut fiber lengths in the increment core (as measured by microscopy), 
simulated under the assumption that the true lengths of those fibers that at least partially appear in the increment core follow a generalized gamma distribution with parameters
\verb"b=2.4, d=3.3" and \verb"k=1.5", the radius of the increment core is \verb"r=2.5". The histogram of the data is shown in Fig. \ref{fig-hist-micro}. 

\begin{verbatim}
> data(microscopy)
> hist(microscopy,breaks=20,main="Microscopy data", 
       xlab="Fiber length (mm)")
\end{verbatim}

These data can be analyzed by calling the function \verb"fled" with the argument \verb"data.type=`microscopy'".
\begin{verbatim}
> m1 <- fled(data=microscopy,data.type="microscopy",model="ggamma",r=2.5) 
> m1

Microscopy data (uncut fibers in the core)

Model: Generalized gamma

Model parameters:
 1.366 1.956 3.444 

'-'Loglik = 275.805  n = 300
\end{verbatim}

The short summary of the results is printed above. It gives information about the type of data and model used, gives the values of the estimated parameters in the order
$b,~d,~k$ together with the value of the minus log likelihood (the  optimization criterion) of the final model and the number of observations used. 

The summary command gives a more detailed overview of the results.

\begin{verbatim}
> summary(m1)

Microscopy data (uncut fibers in the core)

Model: Generalized gamma   Method: ML

Model parameters:
           b_fibers d_fibers k_fibers
Estimate     1.3657   1.9560    3.444
Std. Error   0.8223   0.7890    2.257

Summary statistics for FIBER lengths in the standing tree:
              Mean Std.dev. Skewness Kurtosis
Estimate   2.40458  0.68568   0.3263    3.045
Std. Error 0.05712  0.05292   0.2063    0.265


'-'Loglik = 275.805   Sample size: n = 300

Convergence: Successful completion

\end{verbatim}

By supplying the \verb"fled" object \verb"m1" to the \verb"summary" routine, the summary method first gives some general information about the data and model
being estimated. Then, the estimated parameters are summarized. The estimated model parameters are now returned with the corresponding standard errors. 
The table of summary statistics for fiber lengths in the standing tree is illustrated next. The first row of this table gives the estimates of the expected value of the fiber length, 
its standard deviation, skewness and kurtosis. The standard errors of the mentioned statistics are illustrated in the second row. The information about the convergence of the 
optimization algorithms is printed in the last line.

The estimated density functions can be visualized by using the \verb"plot" routine. 
\begin{verbatim}
> plot(m1)
\end{verbatim}
\begin{figure}[H]\centering
           \includegraphics[height=8cm,width=10cm]{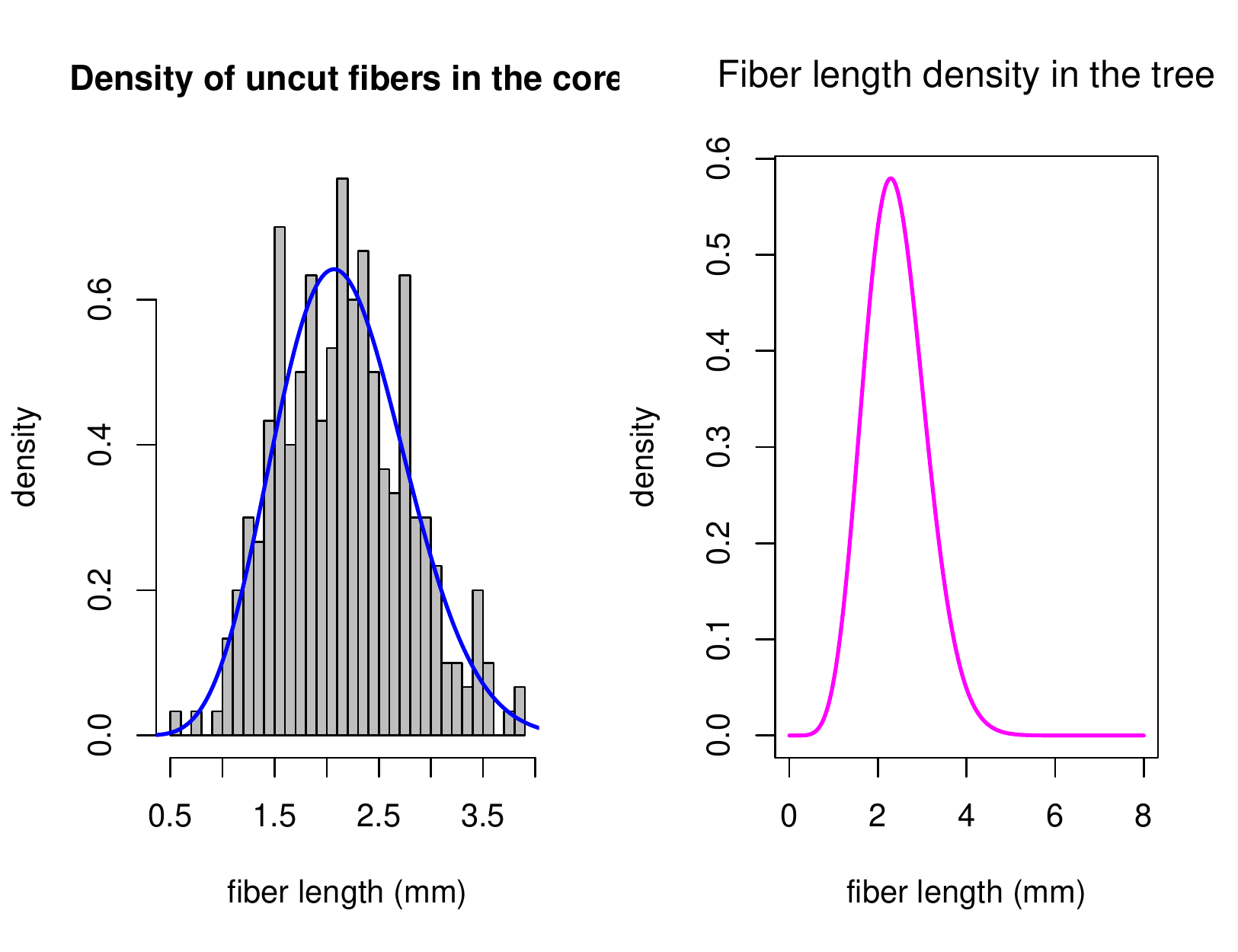}
    \caption{Estimated density functions of model m1.}\label{fig-m1}       
\end{figure}

The left panel of Fig. \ref{fig-m1} shows the estimated density of the uncut fiber lengths in the increment core and the histogram of the given microscopy data.
The estimated density of the fiber lengths in the standing tree is illustrated in the right panel. 

We can also analyze the microscopy data under the assumption that the underlying density function of the true fiber lengths that at least partially appear in the increment core follows a log normal density, using the following code.
\begin{verbatim}
> m2 <- fled(data=dat,data.type="microscopy",model="lognorm",r=2.5)
> plot(m2)
\end{verbatim}
\begin{figure}[H]
  \centering
           \includegraphics[height=8cm,width=10cm]{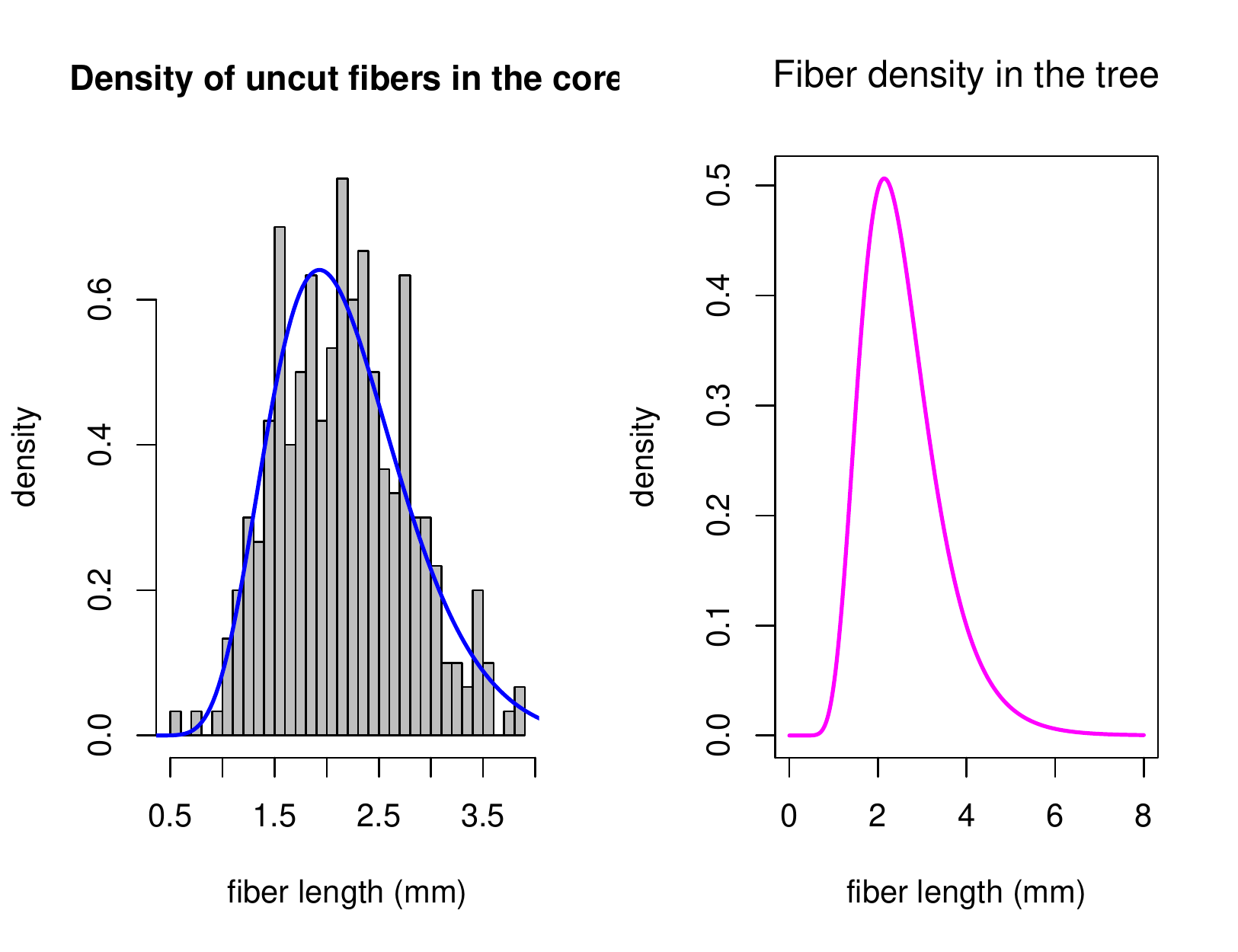}
  \caption{Estimated density plots of model m2.}\label{fig-m2}           
\end{figure}

The resulted plots are shown in Fig. \ref{fig-m2}. Comparing the plots of the density functions on uncut fibers in the increment core (left panels of Figures \ref{fig-m1} and \ref{fig-m2}) for the two considered models, \verb"m1" and
\verb"m2", we can see that the generalized gamma based method better fits the histogram than the log normal one. A summary of various parameter estimates is given by

\begin{verbatim}
> summary(m2)

Microscopy data (uncut fibers in the core)

Model: Log normal   Method: ML

Model parameters:
           mu_fibers sig_fibers
Estimate     0.92869      0.351
Std. Error   0.03575      0.020

Summary statistics for FIBER lengths in the standing tree:
              Mean Std.dev. Skewness Kurtosis
Estimate   2.56329   0.9148  1.11432    5.288
Std. Error 0.09175   0.0821  0.07149    0.304


'-'Loglik = 279.446   Sample size: n = 300

Convergence: Successful completion
\end{verbatim}

 The estimates of the two parameters of the log normal distribution,  $\mu$ and $\sigma,$ are printed in the first row of \verb"Model parameters" table of the 
summary method above, \verb"mu_fibers=0.92869"  and \verb"sig_fibers=0.351", whereas the second row shows the approximate standard errors of the estimates. The table of the summary statistics for fiber lengths in the standing tree 
can be compared with that of the previous model \verb"m1". The log normal based model gives a slightly positively skewed fiber length density in the tree whereas the 
generalized gamma based density function is closer to a symmetrical distribution. The kurtosis of model \verb"m2" is also somewhat higher compared to the kurtosis of model \verb"m1".
Since the `true' values of the parameters of the generalized gamma distribution for the  \verb"microscopy" data are known, we can calculate the corresponding true values for the fiber lengths in the standing tree, which are  \verb"Mean=2.4536", \verb"Std.dev.=0.6723", \verb"Skewness=0.0375" and \verb"Kurtosis=2.7956". The performance using the generalized gamma distribution is better than when using the log normal distribution. This was to be expected since the simulated microscopy data was generated based on a generalized gamma distribution.



 \nocite{*}
 \bibliography{bib-fled-paper}

\begin{appendices}

\section{Maximum likelihood estimation based on OFA data}
The log likelihood function \eqref{LfX} for the OFA data is to be maximised with respect to $\bmth$, under either lognormal or generalized gamma distributional assumptions. This can be done either by direct maximization using a quasi Newton-Raphson method with numerical or analytical derivates, or, if logNormal distributions are assumed, also by the EM algorithm proposed by \cite{svensson2006estimation}. 

\subsection{Generalized gamma mixture model}
Assume the density functions of the true length of fines and fibres in the tree that at least partially appear in the increment core follow generalized gamma densities with
\begin{equation}
\label{GGDfn}
   f_{Y_\fn}(y;\tilde{\bmth}_\fn)= d_1 b_1^{-d_1 k_1}y^{d_1k_1-1}\frac{\exp\left\{-(y/b_1)^{d_1}\right\}}{\Gamma(k_1)}, \;\; y>0,
 \end{equation}
\begin{equation}
\label{GGDfb}
   f_{Y_\fb}(y;\tilde{\bmth}_\fb)= d_2 b_2^{-d_2 k_2}y^{d_2k_2-1}\frac{\exp\left\{-(y/b_2)^{d_2}\right\}}{\Gamma(k_2)}, \;\; y>0,
\end{equation}
where $\tilde{\bmth}_\fn=(b_1, d_1, k_1)^T$,  $\tilde{\bmth}_\fb=(b_2, d_2, k_2)^T$, with all parameters in $\tilde{\bmth}=(\varepsilon,b_1, d_1, k_1,b_2, d_2, k_2)^T$ being positive real-valued numbers.
To ensure positiveness of the six parameters of the generalized gamma distributions and to impose the $[0,1]$ interval restriction on $\varepsilon,$ the following transformations of the  parameters
are considered,
\begin{eqnarray*}
   \varepsilon= \frac{\exp(\theta_1)}{1+\exp(\theta_1)}, ~~~ b_1=\exp(\theta_2), ~~d_1=\exp(\theta_3), ~~k_1=\exp(\theta_4), \\
  b_2=\exp(\theta_5), ~~d_2=\exp(\theta_6), ~~k_2=\exp(\theta_7).
\end{eqnarray*}

The log likelihood function \eqref{LfX}  is now optimized with respect to
\begin{eqnarray}
\label{GGDthetafn}
   \theta_1=\log\frac{\varepsilon}{1-\varepsilon}, ~~ \theta_2=\log(b_1), ~~ \theta_3=\log(d_1), ~~ \theta_4=\log(k_1), \\
   \theta_5=\log(b_2), ~~ \theta_6=\log(d_2), ~~ \theta_7=\log(k_2), \label{GGDthetafb}
\end{eqnarray}

with $\bmth_{\fn}=(\theta_2,\theta_3, \theta_4)^T$, and $\bmth_{\fb}=(\theta_5,\theta_6, \theta_7)^T$, yielding $\bmth=(\theta_1,\bmth_{\fn},\bmth_{\fb})^T$. The optimization can be performed using, e.g., a quasi-Newton's method. Such iterative algorithms need starting values, and may also benefit from knowing the analytic gradient. Below we suggest feasible starting values and also derive the gradient and the Hessian of the log likelihood \eqref{LfX} that may be used in the optimization algorithm.

\subsubsection{Initialization of mixture model parameters}

Starting values for the maximization algorithm is found by solving a simpler maximization problem, assuming that all cells in the increment core are uncut. The distribution of $X$ would then be a mixture of two generalized gamma distributions,
\begin{equation}\label{f0}
   f^{[0]}_X(x;\bmth)=\varepsilon f_{Y_\fn}(x;\bmth_\fn)+(1-\varepsilon)f_{Y_\fb}(x;\bmth_\fb),
\end{equation}
The initial values of the seven model parameters
can then be obtained by maximizing the following log likelihood function.
\begin{equation}\label{llx}
   l^{[0]}(\bfx; \bmth)=\sum\limits_{i=1}^n \log f^{[0]}_X(x_i;\bmth),
\end{equation}

The optimization is performed using the built-in R function \verb"optim()"\citep{R2020}. The default method is set to \verb"L-BFGS-B", a modification of the BFGS quasi-Newton method. If initial values for the parameters to be optimized over are not supplied by a user, then those are set as $\bm\theta^{[0]}=\left(0,\log(.01),\log(.1),\log(10),\log(2),\log(2),\log(2)\right)$. To obtain the gradient of the log likelihood function  (\ref{llx}) used with a quasi-Newton method, we first get the derivatives of the generalized gamma density with respect to its three parameters,
$b,d$ and $k$ and then of the mixture density (\ref{f0}) with respect to all seven transformed parameters. We have, 
\begin{eqnarray}\label{dgg.b}
  \frac{\partial f_{Y_j}(x;\bmth)}{\partial b}&=&d^2b^{-dk-1}x^{dk-1}\frac{\exp\{-(x/b)^d\}}{\Gamma(k)}\left[-k +\left(\frac{x}{b}\right)^d \right] \\ \nonumber
  \frac{\partial f_{Y_j}(x;\bmth)}{\partial d}&=&b^{-dk}x^{dk-1}\frac{\exp\{-(x/b)^d\}}{\Gamma(k)}\left[1 +d\log\left(\frac{x}{b}\right)\left\{k-\left(\frac{x}{b}\right)^d \right\} \right] \\ \nonumber
  \frac{\partial f_{Y_j}(x;\bmth)}{\partial k}&=& db^{-dk}x^{dk-1}\frac{\exp\{-(x/b)^d\}}{\Gamma(k)}\left[d\log\left(\frac{x}{b}\right) -\frac{\Gamma\prime(k)}{\Gamma(k)} \right]
\end{eqnarray}
where $f_{Y_j}(x;\bmth)$ is the density function of a generalized gamma distribution.
Since
$$
  \frac{\textrm{d}\varepsilon}{\textrm{d}\theta_1}= \varepsilon-\varepsilon^2 ~~\textrm{and} ~~\frac{\textrm{d}b_1}{\textrm{d}\theta_2}=\exp(\theta_2)=b_1, ~~\frac{\textrm{d}d_1}{\textrm{d}\theta_3}=\exp(\theta_3)=d_1, ~~
  \frac{\textrm{d}k_1}{\textrm{d}\theta_4}=\exp(\theta_4)=k_1,
$$
and similarly for the three other parameters of the fiber length density, the derivatives of the mixture density (\ref{f0}) will be
\begin{eqnarray}\label{gradEasy}
   \frac{\partial f^{[0]}_X(x;\bmth)}{\partial\theta_1}&=& (\varepsilon-\varepsilon^2)\left\{f_{Y_\fn}(x;\bmth_\fn) - f_{Y_\fb}(x;\bmth_\fb)\right \} \\ \nonumber
   \frac{\partial f^{[0]}_X(x;\bmth)}{\partial\theta_2}&=& b_1  \frac{\partial f_{Y_\fn}(x;\bmth_\fn)}{\partial b_1}, ~~
   \frac{\partial f^{[0]}_X(x;\bmth)}{\partial\theta_3}= d_1  \frac{\partial f_{Y_\fn}(x;\bmth_\fn)}{\partial d_1}, ~~\\ \nonumber
   \frac{\partial f^{[0]}_X(x;\bmth)}{\partial\theta_4}&=& k_1  \frac{\partial f_{Y_\fn}(x;\bmth_\fn)}{\partial k_1}, ~~ \\ \nonumber
   \frac{\partial f^{[0]}_X(x;\bmth)}{\partial\theta_5}&=& b_2  \frac{\partial f_{Y_\fb}(x;\bmth_\fb)}{\partial b_2}, ~~
   \frac{\partial f^{[0]}_X(x;\bmth)}{\partial\theta_6}= d_2  \frac{\partial f_{Y_\fb}(x;\bmth_\fb)}{\partial d_2}, ~~\\ \nonumber
   \frac{\partial f^{[0]}_X(x;\bmth)}{\partial\theta_7}&=& k_2  \frac{\partial f_{Y_\fb}(x;\bmth_\fb)}{\partial k_2}.
\end{eqnarray}
The gradient of the log likelihood (\ref{llx}) is thus $\nabla l^{[0]}(\bfx; \bmth)=\{\partial l^{[0]}(\bfx; \bmth)/{\partial\theta_j}\}_{j=1}^n,$ where
$$
  \frac{\partial l^{[0]}(\bfx; \bmth)}{\partial\theta_j} = \sum_{i}\frac{1}{f^{[0]}_X(x_i;\bmth)} \frac{\partial f^{[0]}_X(x_i;\bmth)}{\partial\theta_j}.
$$

\subsubsection{Derivatives of the full mixture model}

We now obtain the gradient of the log likelihood function of the full mixture model \eqref{LfX} needed for the quasi-Newton steps. From \eqref{pdfX}, Property 3, and (\ref{dgg.b}), straightforward calculations show that the gradient $\nabla l(\bfx; \bmth)=\{\partial l(\bfx; \bmth)/{\partial\theta_j}\}_{j=1}^n$ satisfies
$$
  \frac{\partial l(\bfx; \bmth)}{\partial\theta_j} = \sum_{i}\frac{1}{f_X(x_i;\bmth)} \frac{\partial f_X(x_i;\bmth)}{\partial\theta_j},
$$
with 
\begin{eqnarray}\label{grad}
   \frac{\partial f_X(x;\bmth)}{\partial\theta_1}&=& (\varepsilon-\varepsilon^2)\left\{f_{X_\fn}(x;\bmth_\fn) - f_{X_\fb}(x;\bmth_\fb)\right \} \\ \nonumber
    \frac{\partial f_X(x;\bmth)}{\partial\theta_2}&=& \varepsilon b_1\left[ p_{\textrm{uc}}(x)\frac{\partial f_{Y_\fn}(x;\bmth_\fn)}{\partial\ b_1}
       +\int_{x}^{\infty}f_{X|Y}(x|y)\frac{\partial f_{Y_\fn}(y;\bmth_\fn)}{\partial b_1}dy \right] \\ \nonumber
    \frac{\partial f_X(x;\bmth)}{\partial\theta_3}&=& \varepsilon d_1\left[ p_{\textrm{uc}}(x)\frac{\partial f_{Y_\fn}(x;\bmth_\fn)}{\partial\ d_1}
       +\int_{x}^{\infty}f_{X|Y}(x|y)\frac{\partial f_{Y_\fn}(y;\bmth_\fn)}{\partial d_1}dy \right] \\ \nonumber
    \frac{\partial f_X(x;\bmth)}{\partial\theta_4}&=& \varepsilon k_1\left[ p_{\textrm{uc}}(x)\frac{\partial f_{Y_\fn}(x;\bmth_\fn)}{\partial\ k_1}
       +\int_{x}^{\infty}f_{X|Y}(x|y)\frac{\partial f_{Y_\fn}(y;\bmth_\fn)}{\partial k_1}dy \right] \\ \nonumber
    \frac{\partial f_X(x;\bmth)}{\partial\theta_5}&=& (1-\varepsilon) b_2\left[ p_{\textrm{uc}}(x)\frac{\partial f_{Y_\fb}(x;\bmth_\fb)}{\partial\ b_2}
       +\int_{x}^{\infty}f_{X|Y}(x|y)\frac{\partial f_{Y_\fb}(y;\bmth_\fb)}{\partial b_2}dy \right] \\ \nonumber
    \frac{\partial f_X(x;\bmth)}{\partial\theta_6}&=& (1-\varepsilon) d_2\left[ p_{\textrm{uc}}(x)\frac{\partial f_{Y_\fb}(x;\bmth_\fb)}{\partial\ d_2}
       +\int_{x}^{\infty}f_{X|Y}(x|y)\frac{\partial f_{Y_\fb}(y;\bmth_\fb)}{\partial d_2}dy \right] \\ \nonumber
    \frac{\partial f_X(x;\bmth)}{\partial\theta_7}&=& (1-\varepsilon) k_2\left[ p_{\textrm{uc}}(x)\frac{\partial f_{Y_\fb}(x;\bmth_\fb)}{\partial\ k_2}
       +\int_{x}^{\infty}f_{X|Y}(x|y)\frac{\partial f_{Y_\fb}(y;\bmth_\fb)}{\partial k_2}dy \right].\ \nonumber
\end{eqnarray}

In the above equations we have assumed that it is allowed to change the order of differentiation and integration. 

\subsubsection{Standard errors of the parameter estimates and the Hessian.}\label{secHes}

Under some regularity conditions, the covariance matrix of parameters estimates, $\hat{\bmth}$, can be
approximated by the inverse of the negative observed Hessian, 
$Var(\hat{\bmth})\approx \left(-H \right)^{-1}_{\bmth=\hat{\bmth}},$ where $H=\left\{ \frac{\partial^2 l(\bfx; \bmth)}{\partial\theta_j\partial\theta_i}\right\}.$ Straightforward calculations show that 

{\small
\begin{eqnarray*}
    \frac{\partial^2 l(\bfx; \bmth)}{\partial\theta_1^2}&=&\sum_i\frac{\partial}{\partial\theta_1}\left[ \frac{1}{f_X(x_i;\bmth)}\frac{\partial f_X(x_i;\bmth)}{\partial\theta_1}\right]=-\sum_i\left(\frac{\partial l_i}{\partial\theta_1} \right)^2 +\frac{1-e^{\theta_1}}{1+e^{\theta_1}}
    \sum_i \frac{\partial l_i}{\partial\theta_1} \\ \nonumber
    \frac{\partial^2 l(\bfx; \bmth)}{\partial\theta_1\partial\theta_{(2:4)}}&=&\sum_i\frac{\partial}{\partial\theta_{(2:4)}}\left[ \frac{1}{f_X(x_i;\bmth)}\frac{\partial f_X(x_i;\bmth)}{\partial\theta_1}\right]=\sum_i\left\{-\frac{\partial l_i}{\partial\theta_1}\cdot\frac{\partial l_i}{\partial\theta_{(2:4)}} +
       \frac{1}{1+e^{\theta_1}}\frac{\partial l_i}{\partial\theta_{(2:4)}} \right\} \\ \nonumber
    \frac{\partial^2 l(\bfx; \bmth)}{\partial\theta_1\partial\theta_{(5:7)}}&=& -\sum_i\left\{\frac{\partial l_i}{\partial\theta_1}\cdot\frac{\partial l_i}{\partial\theta_{(5:7)}} + \frac{1}{1+e^{\theta_1}}\frac{\partial l_i}{\partial\theta_{(5:7)}} \right\} \\ \nonumber
    \frac{\partial^2 l(\bfx; \bmth)}{\partial\theta_2^2}&=&\sum_i\left\{-\left(\frac{\partial l_i}{\partial\theta_2} \right)^2 +
              \frac{\varepsilon}{f_X(x_i;\bmth)}\frac{\partial^2 f_{X_\fn}(x_i;\bmth_\fn) }{\partial\theta_2^2} \right\} \\ \nonumber
     \frac{\partial^2 l(\bfx; \bmth)}{\partial\theta_2\partial\theta_{(3,4)}}&=& \sum_i\left\{-\frac{\partial l_i}{\partial\theta_2}\cdot \frac{\partial l_i}{\partial\theta_{(3,4)}} +
              \frac{\varepsilon}{f_X(x_i;\bmth)}\frac{\partial^2 f_{X_\fn}(x_i;\bmth_\fn) }{\partial\theta_2\partial\theta_{(3,4)}} \right\} \\ \nonumber
\end{eqnarray*}
}

{\small
\begin{eqnarray*}
    \frac{\partial^2 l(\bfx; \bmth)}{\partial\theta_2\partial\theta_{(5:7)}}&=& -\sum_i\left(\frac{\partial l_i}{\partial\theta_2}\cdot \frac{\partial l_i}{\partial\theta_{(5:7)}} \right) \\ \nonumber
     \frac{\partial^2 l(\bfx; \bmth)}{\partial\theta_3^2}&=&\sum_i\left\{-\left(\frac{\partial l_i}{\partial\theta_3} \right)^2 +
              \frac{\varepsilon}{f_X(x_i;\bmth)}\frac{\partial^2 f_{X_\fn}(x_i;\bmth_\fn) }{\partial\theta_3^2} \right\} \\ \nonumber
      \frac{\partial^2 l(\bfx; \bmth)}{\partial\theta_3\partial\theta_4}&=& \sum_i\left\{-\frac{\partial l_i}{\partial\theta_3}\cdot \frac{\partial l_i}{\partial\theta_4} +
              \frac{\varepsilon}{f_X(x_i;\bmth)}\frac{\partial^2 f_{X_\fn}(x_i;\bmth_\fn) }{\partial\theta_3\partial\theta_4} \right\} \\ \nonumber
     \frac{\partial^2 l(\bfx; \bmth)}{\partial\theta_3\partial\theta_{(5:7)}}&=& -\sum_i\left(\frac{\partial l_i}{\partial\theta_3}\cdot \frac{\partial l_i}{\partial\theta_{(5:7)}} \right). \\ \nonumber
\end{eqnarray*}
}
Here $l_i = \log f_X(x_i;\bmth)$ and $\partial\theta_{(a,b)}$ means any of $\partial\theta_{a}, \partial\theta_{a+1}, \ldots , \partial\theta_{b}$. The rest of the second order partial derivatives of the log likelihood function  \eqref{LfX}, relating to the fibre lengths,   can be found in a similar way. The second order partial derivatives of the fine length density are given by
{\small
$$
  \frac{\partial^2 f_{X_\fn}(x;\bmth_\fn) }{\partial\theta_k\partial\theta_j}=  p_{\textrm{uc}}(x)\frac{\partial^2 f_{Y_\fn}(x;\bmth_\fn)}{\partial\theta_k\partial\theta_j}
       +\int_{x}^{\infty}f_{X|Y}(x|y)\frac{\partial^2 f_{Y_\fn}(y;\bmth_\fn)}{\partial\theta_k\partial\theta_j}dy,   ~~ k,j=2,3,4
$$
}
and similarly for the second order partial derivatives of the fiber length density $f_{X_\fb}(x;\bmth_\fb)$. Here we have assumed that it is allowed to change the order of differentiation and integration. With the notation
$c_1=\left(\frac{y}{b} \right)^d,$ $c_2=1+(k-c_1)\log(c_1)$ and $c_3=\log(c_1)-\Psi(k),$ the second order partial derivatives of the generalized gamma density are given by

{\small
\begin{eqnarray}\label{d2fy}
  \frac{\partial^2 f_{Y_\fn}(x;\bmth_\fn)}{\partial\theta_2^2}&= &\frac{1}{\Gamma (k)}b^{-dk}d^3y^{dk-1}\exp(-c_1)\left[ (c_1-k)^2 -  c_1\right] \\ \nonumber
  \frac{\partial^2 f_{Y_\fn}(x;\bmth_\fn)}{\partial\theta_2\partial\theta_3}&= & \frac{1}{\Gamma (k)}d^2b^{-dk}y^{dk-1}\exp(-c_1) \left[-c_1 (c_1 - k)\log(c_1)+ \right. \\ \nonumber
     && \left. c_1\log(c_1) + k(c_1-k)\log(c_1) +2(c_1 -k) \right]  \\ \nonumber
  \frac{\partial^2 f_{Y_\fn}(x;\bmth_\fn)}{\partial\theta_2\partial\theta_3}&= & \frac{1}{\Gamma (k)}
     kd^2b^{-dk}y^{dk-1}\exp(-c_1)\left[ \left(c_1 -k\right)\log(c_1)-  \Psi(k)(c_1 -k) -1 \right] \\ \nonumber
   \frac{\partial^2 f_{Y_\fn}(x;\bmth_\fn)}{\partial\theta_3^2}&= & \frac{1}{\Gamma (k)}
     db^{-dk}y^{dk-1}\exp(-c_1)  \left[ 2c_2 -1 +\log(c_1)\left\{kc_2 - c_1\log(c_1) - c_1c_2 \right\}  \right]
      \\ \nonumber
  \frac{\partial^2 f_{Y_\fn}(x;\bmth_\fn)}{\partial\theta_3\partial\theta_4}&= & \frac{1}{\Gamma (k)}
      kdb^{-dk}y^{dk-1}\exp(-c_1) \left[  c_2\log(c_1) - c_2\Psi(k)  + \log(c_1)  \right]
      \\ \nonumber
  \frac{\partial^2 f_{Y_\fn}(x;\bmth_\fn)}{\partial\theta_4^2}&= & \frac{1}{\Gamma (k)}
     dkb^{-dk}y^{dk-1}\exp(-c_1)  \left[ kc_3\log(c_1) - kc_3\Psi(k) + c_3 - k\Psi_1(k) \right],
\end{eqnarray}
}
where $\Psi(k)$ and $\Psi_1(k)$ are digamma and trigamma functions correspondingly.
\\
\\
A simple and direct way to approximate the covariance matrix of the parameter estimates on the original scale, $\hat{\tilde\bmth},$ is to use the delta method, yielding
$$
  V_{\tilde\bmth}=Var(\hat{\tilde\bmth}) = \textrm{diag}\left(\nabla\hat{\tilde\bmth}\right) Var(\hat{\bmth}) \textrm{diag}\left(\nabla\hat{\tilde\bmth} \right),
$$
where $\textrm{diag}\left(\nabla\hat{\tilde\bmth}\right)$ is a diagonal matrix of size seven, with the vector of the first order derivatives of $\tilde{\bmth}$ with respect to $\bmth$
evaluated at $\hat\bmth,$ on the main diagonal. Note that 
$$
  \frac{\partial\tilde{\bmth}}{\partial\bmth}=(\varepsilon-\varepsilon^2, \exp(\theta_2),\exp(\theta_3),\exp(\theta_4),\exp(\theta_5),\exp(\theta_6),\exp(\theta_7))^T.
$$

\subsection{Summary statistics and their standard errors}
The expression for the means of the distribution of fine and fiber lengths in a standing tree is given in Property 1. 
Below we also give the expressions for three more summary statistics for fiber and fine length distributions in the standing tree being the standard deviation, skewness and kurtosis. Plug-in estimates of these quantities together with estimated standard errors for the estimates are also given. 
Below we provide the fine length summary statistics. The summary statistics of the fiber lengths is obtained by simply replacing the index \textit{fines} with \textit{fibers}.

From Property 1 it may be concluded that the $m^{\textrm th}-$ moment of the fine length distribution satisfies
\begin{equation*}
  E_\bmth(W^m_\fn)=\int_{0}^{\infty} \frac{y^m\left\{\pi r +2E_\bmth(W_\fn) \right\}}{\pi r+2y} f_{Y_\fn}(y;\bmth_\fn)dy.
\end{equation*}
The summary statistics of interest can be formulated in terms of such moments as
\begin{eqnarray*}
  \mu_\fn&=&E_\bmth(W_\fn)\\
  \sigma^2_\fn &=&Var_\bmth(W_\fn)  =  E_\bmth(W^2_\fn) - \mu_\fn^2 \\
  \textrm{Skewness}_\fn&=& \frac{1}{\sigma^3_\fn} \left\{ E_\bmth(W^3_\fn) -3\mu_\fn E_\bmth(W^2_\fn)+2\mu_\fn^3    \right\} \\
  \textrm{Kurtosis}_\fn &=& \frac{1}{\sigma^4_\fn} \left\{ E_\bmth(W^4_\fn) -4\mu_\fn E_\bmth(W^3_\fn)+6\mu_\fn^2 \sigma^2_\fn + 3\mu_\fn^4 \right\}
\end{eqnarray*}
Note that, for simplicity, the dependence of $\mu_\fn$ and $\sigma_\fn$ on $\bmth_\fn$ has been omitted. The estimated quantities are found by replacing $\bmth$ by $\hat{\bmth}$ in the above equations, so called plug-in estimates.

  \subsubsection{Standard errors of summary statistics via the delta method }
Let $g(\hat{\bmth})$ denote an estimated summary statistic. Using the delta method the variance of $g(\hat{\bmth})$ can be approximated by
$$
  Var(g(\hat{\bmth}))=\nabla g(\hat{\bmth})^T Var(\hat{\bmth})\nabla g(\hat{\bmth}),
$$
where $\nabla g(\bmth)$ is the gradient of $g(\bmth)$ and $Var(\hat{\bmth})$ was defined in Section \ref{secHes}.
The gradients of the summary statistics may be obtained as follows.

{\small
\begin{eqnarray*}
  \frac{\partial \mu_\fn}{\partial \theta_j} &=& -\frac{1}{2}\left[ \int_{0}^{\infty}\frac{f_{Y_\fn}(y;\bmth_\fn)}{\pi r+2y}dy\right]^{-2}
       \int_{0}^{\infty}\frac{\partial f_{Y_\fn}(y;\bmth_\fn)}{\partial \theta_j} \frac{1}{\pi r+2y}dy, ~~~j=2,3,4 \\
  \frac{\partial \sigma^2_\fn}{\partial \theta_j} &=& \frac{\partial E_\bmth(W^2_\fn) }{\partial \theta_j} -2\mu_\fn\frac{\partial \mu_\fn}{\partial \theta_j} \\
  \frac{\partial E_\bmth(W^m_\fn)}{\partial \theta_j} &=& \int_{0}^{\infty}\frac{y^m}{\pi r +2y}\left\{2f_{Y_\fn}(y;\bmth_\fn)\frac{\partial\mu_\fn}{\partial\theta_j}+
            (\pi r+2\mu_\fn)\frac{\partial f_{Y_\fn}(y;\bmth_\fn)}{\partial\theta_j}  \right\}dy \\
  \frac{\partial \textrm{Skewness}_\fn}{\partial \theta_j} &=& \frac{1}{\sigma_\fn^3}\left\{\frac{\partial E_\bmth(W^3_\fn)}{\partial \theta_j}-3\mu_\fn\frac{\partial E_\bmth(W^2_\fn)}{\partial\theta_j}-3E_\bmth(W^2_\fn)\frac{\partial \mu_\fn}{\partial\theta_j}+6\mu^2_\fn\frac{\partial \mu_\fn}{\partial \theta_j}   \right\}   \\
        & & - \frac{3}{2\sigma_\fn^5}\frac{\partial \sigma^2_\fn}{\partial \theta_j}\left\{ E_\bmth(W^3_\fn)-3\mu_\fn E_\bmth(W^2_\fn) +2\mu_\fn^3 \right\} \\
  \frac{\partial \textrm{Kurt}_\fn}{\partial \theta_j} &=& \frac{1}{\sigma_\fn^4}\left[ \frac{\partial E_\bmth(W^4_\fn)}{\partial \theta_j}
       -4\mu_\fn\frac{\partial E_\bmth(W^3_\fn)}{\partial \theta_j} +6\mu_\fn^2 \frac{\partial \sigma^2_\fn}{\partial \theta_j} \right. \\
      && \left. +4\frac{\partial \mu_\fn}{\partial \theta_j}\left\{3\mu^3_\fn +  3\mu_\fn\sigma^2_\fn -E_\bmth(W^3_\fn) \right\}   \right] \\
     &&  -\frac{2}{\sigma^6_\fn}\frac{\partial \sigma^2_\fn}{\partial \theta_j}\left\{ E_\bmth(W^4_\fn) -4\mu_\fn E_\bmth(W^3_\fn)+6\mu_\fn^2 \sigma^2_\fn + 3\mu_\fn^4 \right\}
\end{eqnarray*}
}
Note that the above derivatives are taken with respect to the three parameters of the fine length distribution, $\theta_2, \theta_3$ and $\theta_4$, and that we have assumed that it is allowed to change the order of differentiation and integration. To find the standard errors of the the proportion of fines in the tree, $\tilde\varepsilon,$ given in (\ref{tilde.eps}), we first obtain the partial derivatives of the expected value of the cell length, $E(W),$ in (\ref{EW}),
using the chain rule, yielding

{\small
\begin{eqnarray*}
     \frac{\partial E(W)}{\partial \theta_1} &=& \frac{(\varepsilon^2-\varepsilon)(\mu_\fb-\mu_\fn)}{2(\varepsilon\mu_\fb+ (1-\varepsilon)\mu_\fn)+\pi r}
         \left[\pi r+\frac{2(2\mu_\fn\mu_\fb +\varepsilon\mu_\fn\pi r+(1-\varepsilon)\mu_\fb\pi r)}{2(\varepsilon\mu_\fb+ (1-\varepsilon)\mu_\fn)+\pi r} \right] \\
     \frac{\partial E(W)}{\partial \theta_{(2:4)}} &=& \frac{\partial E(W)}{\partial \mu_\fn}\frac{\partial\mu_\fn}{\partial \theta_{(2:4)}}, ~~~~
     \frac{\partial E(W)}{\partial \theta_{(5:7)}} = \frac{\partial E(W)}{\partial \mu_\fb}\frac{\partial\mu_\fb}{\partial \theta_{(5:7)}} \\
     \frac{\partial E(W)}{\partial \mu_\fn} &=& \frac{2\mu_\fb +\varepsilon\pi r}{2(\varepsilon\mu_\fb+ (1-\varepsilon)\mu_\fn)+\pi r}
           - \frac{2(1-\varepsilon)\left\{2\mu_\fn\mu_\fb +\varepsilon\mu_\fn\pi r+(1-\varepsilon)\mu_\fb\pi r\right\}}{\left\{2(\varepsilon\mu_\fb+ (1-\varepsilon)\mu_\fn)+\pi r\right\}^2}  \\
     \frac{\partial E(W)}{\partial \mu_\fb} &=& \frac{2\mu_\fn +(1-\varepsilon)\pi r}{2(\varepsilon\mu_\fb+ (1-\varepsilon)\mu_\fn)+\pi r}
           - \frac{2\varepsilon\left\{2\mu_\fn\mu_\fb +\varepsilon\mu_\fn\pi r+(1-\varepsilon)\mu_\fb\pi r\right\}}{\left\{2(\varepsilon\mu_\fb+ (1-\varepsilon)\mu_\fn)+\pi r\right\}^2}.  \\
\end{eqnarray*}
}

Finally,
{\small
\begin{eqnarray*}
  \frac{\partial\tilde\varepsilon}{\partial\theta_1} &=& (\varepsilon-\varepsilon^2)\frac{\pi r +2E(W)}{\pi r+2\mu_\fn} +\frac{2\varepsilon}{\pi r+2\mu_\fn}\frac{\partial E(W)}{\partial \theta_1}  \\
   \frac{\partial\tilde\varepsilon}{\partial\theta_{(2:4)}} &=& \frac{2\varepsilon}{\pi r+2\mu_\fn} \left\{ \frac{\partial E(W)}{\partial \mu_\fn} -
       \frac{\pi r +2E(W)}{\pi r+2\mu_\fn}\frac{\partial \mu_\fn}{\partial \theta_{(2:4)}}   \right\} \\
   \frac{\partial\tilde\varepsilon}{\partial\theta_{(2:4)}} &=&  \frac{2\varepsilon}{\pi r+2\mu_\fn} \frac{\partial E(W)}{\partial \mu_\fb}.
\end{eqnarray*}
}

\subsection{Log normal mixture model}
Assume now that the density functions of the true length of fines and fibres that at least partially appear in the increment core follow the log normal densities
 \begin{equation}
 \label{LogNfn}
   f_{Y_\fn}(y;\tilde{\bmth}_\fn)= \frac{1}{y\sigma_1\sqrt{2\pi}}e^{-\frac{(\log y -\mu_1)^2}{2\sigma^2_1}}, \; \;y>0,
\end{equation}
 \begin{equation}
  \label{LogNfb}
   f_{Y_\fb}(y;\tilde{\bmth}_\fb)= \frac{1}{y\sigma_2\sqrt{2\pi}}e^{-\frac{(\log y -\mu_2)^2}{2\sigma^2_2}}, \; \; y>0,
\end{equation}
where $\tilde{\bmth}_\fn=(\mu_1, \sigma_1)^T$, $\tilde{\bmth}_\fb=(\mu_2, \sigma_2)^T$, and $\sigma_1, \sigma_2>0$. 
To ensure positiveness of $\sigma_1$ and $\sigma_2$ and to impose the $[0,1]$ interval restriction on $\varepsilon,$ the following transformations of the  lognormal parameters are considered,
\begin{eqnarray*}
   \varepsilon= \frac{\exp(\tau)}{1+\exp(\tau)}, ~~~ \sigma_1=\exp(\theta_1), ~~\sigma_2=\exp(\theta_2).
\end{eqnarray*}

The log likelihood function (\ref{LfX}) is  optimized with respect to
\begin{eqnarray}
\label{LogNtheta}
   \bmth=\left(\tau=\log\frac{\varepsilon}{1-\varepsilon}, ~~ \mu_1, ~~ \theta_1=\log(\sigma_1), ~~ \mu_2, ~~ \theta_2=\log(\sigma_2)\right)^T.
\end{eqnarray}

Below we suggest feasible starting values and also derive the gradient of the log likelihood (\ref{LfX}).

\subsubsection{Initialization of mixture model parameters}

Starting values for the maximization algorithm is found by solving a simpler maximization problem. Assuming that all cells in the increment core are uncut, the distribution of $X$ would be a mixture of two log normal distributions of the form (15). The following derivatives are useful for an optimization procedure.
\begin{eqnarray*}
    \frac{\partial f_{Y_j}(x;\bmth)}{\partial\mu}&=& \frac{\log x-\mu}{\sigma^2}f_{Y_j}(x;\bmth) \\
    \frac{\partial f_{Y_j}(x;\bmth)}{\partial\theta}&=& \left\{\frac{(\log x-\mu)^2}{\sigma^2} -1\right\}f_{Y_j}(x;\bmth),
\end{eqnarray*}
 where $f_{Y_j}(x;\bmth)$ is the pdf of the log normal distribution and $\theta=\log(\sigma).$

\begin{equation*}
   l^{[0]}(\bfx; \bmth)= \sum\limits_{i=1}^n \log f^{[0]}_X(x;\bmth)=\sum\limits_{i=1}^n \log\left\{\varepsilon f_{Y_\fn}(x;\bmth_\fn)+(1-\varepsilon)f_{Y_\fb}(x;\bmth_\fb) \right\},
\end{equation*}
\begin{eqnarray*}
    \frac{\partial l^{[0]}}{\partial\tau}&=& (\varepsilon-\varepsilon^2)\sum_i\frac{1}{f^{[0]}_X(x;\bmth)}\left\{f_{Y_\fn}(x;\bmth_\fn)-f_{Y_\fb}(x;\bmth_\fb) \right\}   \\
     \frac{\partial l^{[0]}}{\partial\mu_1}&=& \sum_i\frac{\varepsilon}{f^{[0]}_X(x;\bmth)}\frac{\partial f_{Y_\fn}(x;\bmth_\fn)}{\partial\mu_1}, ~~~~
      \frac{\partial l^{[0]}}{\partial\theta_1}= \sum_i\frac{\varepsilon}{f^{[0]}_X(x;\bmth)}\frac{\partial f_{Y_\fn}(x;\bmth_\fn)}{\partial\theta_1} \\
      \frac{\partial l^{[0]}}{\partial\mu_2}&=& \sum_i\frac{1-\varepsilon}{f^{[0]}_X(x;\bmth)}\frac{\partial f_{Y_\fb}(x;\bmth_\fb)}{\partial\mu_2}, ~~~~
      \frac{\partial l^{[0]}}{\partial\theta_2}= \sum_i\frac{1-\varepsilon}{f^{[0]}_X(x;\bmth)}\frac{\partial f_{Y_\fb}(x;\bmth_\fb)}{\partial\theta_2} \\
\end{eqnarray*}

\subsubsection{Some derivatives of the full log normal mixture model}

The derivatives are similar to those of the generalized gamma mixture model (\ref{grad}).

\begin{eqnarray*}
   \frac{\partial f_X(x;\bmth)}{\partial\tau}&=& (\varepsilon-\varepsilon^2)\left\{f_{X_\fn}(x;\bmth_\fn) - f_{X_\fb}(x;\bmth_\fb)\right \} \\ \nonumber
    \frac{\partial f_X(x;\bmth)}{\partial\mu_1}&=& \varepsilon \left[ p_{\textrm{uc}}(x)\frac{\partial f_{Y_\fn}(x;\bmth_\fn)}{\partial\mu_1}
       +\int_{x}^{\infty}f_{X|Y}(x|y)\frac{\partial f_{Y_\fn}(y;\bmth_\fn)}{\partial\mu_1}dy \right] \\ \nonumber
    \frac{\partial f_X(x;\bmth)}{\partial\theta_1}&=& \varepsilon \left[ p_{\textrm{uc}}(x)\frac{\partial f_{Y_\fn}(x;\bmth_\fn)}{\partial\theta_1}
       +\int_{x}^{\infty}f_{X|Y}(x|y)\frac{\partial f_{Y_\fn}(y;\bmth_\fn)}{\partial\theta_1}dy \right] \\ \nonumber
    \frac{\partial f_X(x;\bmth)}{\partial\mu_2}&=& (1-\varepsilon)\left[ p_{\textrm{uc}}(x)\frac{\partial f_{Y_\fb}(x;\bmth_\fb)}{\partial\mu_2}
       +\int_{x}^{\infty}f_{X|Y}(x|y)\frac{\partial f_{Y_\fb}(y;\bmth_\fb)}{\partial\mu_2}dy \right] \\ \nonumber
    \frac{\partial f_X(x;\bmth)}{\partial\theta_2}&=& (1-\varepsilon)\left[ p_{\textrm{uc}}(x)\frac{\partial f_{Y_\fb}(x;\bmth_\fb)}{\partial\theta_2}
       +\int_{x}^{\infty}f_{X|Y}(x|y)\frac{\partial f_{Y_\fb}(y;\bmth_\fb)}{\partial\theta_2}dy \right] \\ \nonumber
\end{eqnarray*}

\subsection{The log normal mixture Hessian}

\begin{eqnarray*}
    \frac{\partial^2 l(\bfx; \bmth)}{\partial\tau^2}&=& -\sum_i\left(\frac{\partial l_i}{\partial\tau} \right)^2 +\frac{1-e^{\tau}}{1+e^{\tau}}
            \sum_i \frac{\partial l_i}{\partial\tau} \\ 
    \frac{\partial^2 l(\bfx; \bmth)}{\partial\tau\partial\mu_1}&=& \sum_i\left\{-\frac{\partial l_i}{\partial\tau}\cdot\frac{\partial l_i}{\partial\mu_1} +
            \frac{1}{1+e^{\tau}}\frac{\partial l_i}{\partial\mu_1} \right\} \\ 
    \frac{\partial^2 l(\bfx; \bmth)}{\partial\tau\partial\theta_1}&=& \sum_i\left\{-\frac{\partial l_i}{\partial\tau}\cdot\frac{\partial l_i}{\partial\theta_1} +
            \frac{1}{1+e^{\tau}}\frac{\partial l_i}{\partial\theta_1} \right\} \\         
    \frac{\partial^2 l(\bfx; \bmth)}{\partial\tau\partial\mu_2}&=& -\sum_i\left\{\frac{\partial l_i}{\partial\tau}\cdot\frac{\partial l_i}{\partial\mu_2} + 
                      \frac{1}{1+e^{\tau}}\frac{\partial l_i}{\partial\mu_2} \right\} \\
    \frac{\partial^2 l(\bfx; \bmth)}{\partial\tau\partial\theta_2}&=& -\sum_i\left\{\frac{\partial l_i}{\partial\tau}\cdot\frac{\partial l_i}{\partial\theta_2} +
                      \frac{1}{1+e^{\tau}}\frac{\partial l_i}{\partial\theta_2} \right\} \\                  
    \frac{\partial^2 l(\bfx; \bmth)}{\partial\mu_1^2}&=&\sum_i\left\{-\left(\frac{\partial l_i}{\partial\mu_1} \right)^2 +
                 \frac{\varepsilon}{f_X(x_i;\bmth)}\frac{\partial^2 f_{X_\fn}(x_i;\bmth_\fn) }{\partial\mu_1^2} \right\} \\ 
     \frac{\partial^2 l(\bfx; \bmth)}{\partial\mu_1\partial\theta_1}&=& \sum_i\left\{-\frac{\partial l_i}{\partial\mu_1}\cdot \frac{\partial l_i}{\partial\theta_1} +
              \frac{\varepsilon}{f_X(x_i;\bmth)}\frac{\partial^2 f_{X_\fn}(x_i;\bmth_\fn) }{\partial\mu_1\partial\theta_1} \right\} \\ 
     \frac{\partial^2 l(\bfx; \bmth)}{\partial\mu_1\partial\mu_2(\theta_2)}&=& -\sum_i\left(\frac{\partial l_i}{\partial\mu_1}\cdot \frac{\partial l_i}{\partial\mu_2(\theta_2)} \right) \\
     \frac{\partial^2 l(\bfx; \bmth)}{\partial\theta_1^2}&=&\sum_i\left\{-\left(\frac{\partial l_i}{\partial\theta_1} \right)^2 +
              \frac{\varepsilon}{f_X(x_i;\bmth)}\frac{\partial^2 f_{X_\fn}(x_i;\bmth_\fn) }{\partial\theta_1^2} \right\} \\ 
     \frac{\partial^2 l(\bfx; \bmth)}{\partial\theta_1\partial\mu_2(\theta_2)}&=& -\sum_i\left(\frac{\partial l_i}{\partial\theta_1}\cdot \frac{\partial l_i}{\partial\mu_2(\theta_2)} \right) \\                 
      \frac{\partial^2 l(\bfx; \bmth)}{\partial\mu_2^2}&=&\sum_i\left\{-\left(\frac{\partial l_i}{\partial\mu_2} \right)^2 +
                 \frac{1-\varepsilon}{f_X(x_i;\bmth)}\frac{\partial^2 f_{X_\fb}(x_i;\bmth_\fb) }{\partial\mu_2^2} \right\} \\    
      \frac{\partial^2 l(\bfx; \bmth)}{\partial\mu_2\partial\theta_2}&=& \sum_i\left\{-\frac{\partial l_i}{\partial\mu_2}\cdot \frac{\partial l_i}{\partial\theta_2} +
              \frac{1-\varepsilon}{f_X(x_i;\bmth)}\frac{\partial^2 f_{X_\fb}(x_i;\bmth_\fb) }{\partial\mu_2\partial\theta_2} \right\} \\  
      \frac{\partial^2 l(\bfx; \bmth)}{\partial\theta_2^2}&=&\sum_i\left\{-\left(\frac{\partial l_i}{\partial\theta_2} \right)^2 +
              \frac{1-\varepsilon}{f_X(x_i;\bmth)}\frac{\partial^2 f_{X_\fb}(x_i;\bmth_\fb) }{\partial\theta_2^2} \right\}, \\         
\end{eqnarray*}
where
{\small
$$
  \frac{\partial^2 f_{X_\fn}(x;\bmth_\fn) }{\partial\theta_k\partial\theta_j}=  p_{\textrm{uc}}(x)\frac{\partial^2 f_{Y_\fn}(x;\bmth_\fn)}{\partial\theta_k\partial\theta_j}
       +\int_{x}^{\infty}f_{X|Y}(x|y)\frac{\partial^2 f_{Y_\fn}(y;\bmth_\fn)}{\partial\theta_k\partial\theta_j}dy,   ~~ k,j=2,3,4
$$
}
and similar for the second order partial derivatives of the fiber length density. We now need the second order partial derivatives of the log normal density.

{\small
\begin{eqnarray*}
  \frac{\partial^2 f_{Y_\fn}(x;\bmth_\fn)}{\partial\mu_1^2}&= & \frac{(\log x -\mu_1)}{\sigma_1^2} \frac{\partial f_{Y_\fn}(x;\bmth_\fn)}{\partial\mu_1}-\frac{f_{Y_\fn}(x;\bmth_\fn)}{\sigma_1^2} \\
  \frac{\partial^2 f_{Y_\fn}(x;\bmth_\fn)}{\partial\theta_1^2}&= & \left\{\frac{(\log x -\mu_1)^2}{\sigma_1^2}-3 \right\}\frac{\partial f_{Y_\fn}(x;\bmth_\fn)}{\partial\theta_1}-2f_{Y_\fn}(x;\bmth_\fn) \\
  \frac{\partial^2 f_{Y_\fn}(x;\bmth_\fn)}{\partial\mu_1\partial\theta_1}&= & \frac{(\log x -\mu_1)}{\sigma_1^2} \frac{\partial f_{Y_\fn}(x;\bmth_\fn)}{\partial\theta_1}-2\frac{\partial f_{Y_\fn}(x;\bmth_\fn)}{\partial\mu_1} 
\end{eqnarray*}
}

Finally,
$$
  V_{\tilde\bmth}=Var(\hat{\tilde\bmth}) = \textrm{diag}\left(\nabla\hat{\tilde\bmth}\right) Var(\hat{\bmth}) \textrm{diag}\left(\nabla\hat{\tilde\bmth} \right),
$$
where $\textrm{diag}\left(\nabla\hat{\tilde\bmth}\right)$ is a diagonal matrix of size five, with the vector of the first order derivatives of $\tilde{\bmth}$ with respect to $\bmth$
evaluated at $\hat\bmth,$ on the main diagonal.
$$
  \frac{\partial\tilde{\bmth}}{\partial\bmth}=(\varepsilon-\varepsilon^2, 1,\exp(\theta_1),1,\exp(\theta_2))^T.
$$

\section{Some derivatives needed when dealing with Microscopy data}
 The first derivatives of log generalized gamma density $f_{Y}(y;\bmth).$
$$
  \frac{\partial \log f_Y}{\partial\theta_1}= d\left\{ \left(\frac{y}{b}\right)^d -k \right\}
$$
$$
  \frac{\partial \log f_Y}{\partial\theta_2}= 1+d\log\left(\frac{y}{b}\right) \left\{k- \left(\frac{y}{b}\right)^d \right\}
$$
$$
  \frac{\partial \log f_Y}{\partial\theta_3}= k\left\{ d\log\left(\frac{y}{b}\right) -\frac{\Gamma'(k)}{\Gamma(k)} \right\}
$$

The log likelihood of the observed microscopy sample can be written as
$$
  L({\bf x;\bmth})=\prod_{i=1}^n f_M(x_i, \bmth)=\prod_{i=1}^n \frac{f_Y(x_i,\bmth)p_{uc}(x_i,r)}{\int_0^{2r}f_Y(c,\bmth)p_{uc}(c,r)dc}.
$$
Then the gradient of the log likelihood function can be found with the following elements.
$$
   \frac{\partial l_i({x_i;\bmth})}{\partial\theta_j}=\frac{1}{f_Y(x_i,\bmth)}\frac{\partial f_Y(x_i,\bmth)}{\partial\theta_j} - \frac{1}{\int_0^{2r}f_Y(c,\bmth)p_{uc}(c,r)dc}
       \int\limits_0^{2r}\frac{\partial f_Y(c,\bmth)}{\partial\theta_j}p_{uc}(c,r)dc
$$
Finally, below is an expression for the elements of the Hessian matrix.
\begin{eqnarray*}
     \frac{\partial^2 l_i({x_i;\bmth})}{\partial\theta_j\partial\theta_k} & = &  -\frac{1}{f^2_Y(x_i,\bmth)}\frac{\partial f_Y(x_i,\bmth)}{\partial\theta_j}\frac{\partial f_Y(x_i,\bmth)}{\partial\theta_k}
     + \frac{1}{f_Y(x_i,\bmth)}\frac{\partial^2 f_Y(x_i,\bmth)}{\partial\theta_j\partial\theta_k} \\ 
     &+ &   \frac{1}{\left(\int\limits_0^{2r}f_Y(c,\bmth)p_{uc}(c,r)dc\right)^2} \int\limits_0^{2r}\frac{\partial f_Y(c,\bmth)}{\partial\theta_j}p_{uc}(c,r)dc \int\limits_0^{2r}\frac{\partial f_Y(c,\bmth)}{\partial\theta_k}p_{uc}(c,r)dc
     \\
     & - &  \frac{1}{\int_0^{2r}f_Y(c,\bmth)p_{uc}(c,r)dc}\int\limits_0^{2r}\frac{\partial^2 f_Y(c,\bmth)}{\partial\theta_j\partial\theta_k}p_{uc}(c,r)dc
\end{eqnarray*}

The gradients of the log normal density can be obtained in a similar way.

\end{appendices}

\end{document}